\documentclass[twocolumn]{aastex631}
\usepackage{color} 

\usepackage{epsfig}
\usepackage{subfigure}
\usepackage{graphicx}
\usepackage{amsmath}
\usepackage{natbib}
\newcommand{\pa}{\partial}
\newcommand{\mb}{\boldsymbol}
\newcommand{\mc}{\mathcal}

\newcommand{\bgeq}{\begin{equation}}
\newcommand{\edeq}{\end{equation}}

\defcitealias{Bai_etal19}{BOPS19}

\shorttitle{CR Transport Coefficients from Streaming Box Simulations}
\shortauthors{X.-N. Bai}

\begin{document}


\title{Towards First-principle Characterization of Cosmic-ray Transport Coefficients from Multi-scale Kinetic Simulations}


\author{Xue-Ning Bai}
\affil{Institute for Advanced Study, Tsinghua University, Beijing 100084, China}
\affil{Department of Astronomy, Tsinghua University, Beijing 100084, China}
\email{xbai@tsinghua.edu.cn}




\begin{abstract}
A major uncertainty in understanding the transport and feedback of cosmic-rays (CRs) within and beyond
our Galaxy lies in the unknown CR scattering rates, which are primarily
determined by wave-particle interaction at microscopic gyro-resonant scales. The source of the waves
for the bulk CR population is believed to be self-driven by the CR streaming instability (CRSI), resulting
from the streaming of CRs downward a CR pressure gradient.
While a balance between driving by the CRSI and wave damping is expected to determine wave
amplitudes and hence the CR scattering rates, the problem involves significant scale separation with
substantial ambiguities based on quasi-linear theory (QLT).
Here we propose a novel ``streaming box" framework to study the CRSI with an imposed CR pressure gradient,
enabling first-principle measurement of the CR scattering rates as a function of environmental parameters.
By employing the magnetohydrodynamic-particle-in-cell (MHD-PIC) method with ion-neutral damping,
we conduct a series of simulations with different resolutions and CR pressure gradients and precisely
measure the resulting CR scattering rates in steady state. The measured rates show scalings consistent
with QLT, but with a normalization smaller by a factor of several than typical estimates based on
single-fluid treatment of CRs.
A momentum-by-momentum treatment provides better estimates when integrated over momentum, but is also subject substantial deviations especially at small momentum. 
Our framework thus opens up the path towards providing comprehensive subgrid physics for
macroscopic studies of CR transport and feedback in broad astrophysical contexts.
\end{abstract}



\section{Introduction}\label{sec:intro}

Cosmic-rays (CRs) are (trans-)relativistic charged particles traveling through space. They are
generated in energetic astrophysical sources, especially shocks from supernova remnants
via diffusive shock acceleration
\citep[e.g.]{Bell78,BlandfordOstriker78,Drury83}. Being energetic particles, CRs
naturally tend to escape from acceleration sites. Individual CR particles travel along while
gyrating around magnetic field lines, and in the meantime are scattered by waves and turbulence.
The intensity of the waves and/or level of turbulence determine how efficiently CRs propagate
through the background medium, a process known as CR transport. In the Galaxy, the bulk
CR energy density is dominated by protons with $\sim$GeV energies that are primarily generated from the Galactic
disk. They are transported outward, and eventually escape from the Galaxy
over a few million years \citep{GinzburgSyrovatskii64,Strong_etal07}. Balancing CR production
with escape (and to a lesser extent, energy losses), the typical CR energy density found in the
interstellar medium (ISM) is of the order $\sim1$ eV cm$^{-3}$, corresponding to a number density of
$n_{\rm CR}\sim10^{-9}$ cm$^{-3}$ \citep{Grenier_etal15}.

In the interstellar medium (ISM), CRs are not simply test particles that respond passively to
background waves and turbulence. Rather, they are dynamically important with energy
density comparable to or exceeding that of other components of the ISM such as thermal gas and
magnetic field. As such, the backreaction from the CRs to their host galaxy
leads to a number of significant physical consequences, known as CR feedback
\citep[see e.g. the reviews of][]{Ferriere01,Zweibel17,NaabOstriker17}.
The coupling between CRs and background thermal gas is mediated by magnetic fields.
Besides pressure support, wave-particle interaction allows CRs to exchange momentum
and energy with the waves, and hence the background gas itself, leading to heating and
momentum deposition in the ISM, potentially driving galactic winds
\citep[e.g.,][]{Ipavich75,Breitschwerdt_etal91,Zirakashvili_etal96,Everett_etal08,MaoOstriker18,Hopkins_etal21a,Quataert_etal21b,HuangDavis22}.

The most complex but fascinating fact about CR transport and feedback is that the waves that are crucially
responsible are considered to be self-generated due to the CR streaming instability 
\citep[CRSI;][]{KulsrudPearce69,Skilling71,Wentzel74}\footnote{This
is the case for the bulk CR population, whereas scattering of more energetic CRs (above a few
hundred GeV, \citealp{Blasi_etal12,Aloisio_etal15}) are likely dominated by extrinsic turbulence in the ISM
\citep[e.g.,][]{Jokipii66,SchlickeiserMiller98,YanLazarian02}.}: when the bulk CRs drift (a.k.a. streaming)
relative to background gas at a speed $v_D$ that is faster than the Alfv\'en speed $v_A$, Alfv\'en waves
become unstable, and grow at the cost of the free energy from CR streaming. The underlying physical
mechanism of this so-called CR self-confinement
involves the gyro-resonance: particles exchange energy and momentum only to Alfv\'en waves
whose wavelengths are resonant with its gyro-orbit. The growth rate of the CRSI is
of the order $\Omega_c(v_D-v_A)(n_{\rm CR}/n_0)$, where $\Omega_c$ is the cyclotron frequency, and
$n_{\rm CR}$, $n_0$ are number densities of the CRs and background ion plasma.
Usually, CR streaming is naturally fed by the CR sources as CRs escape down a CR pressure gradient.
The Alfv\'en waves generated by the CRSI will scatter the CRs, and are expected to isotropize
the CR distribution in the wave frame, thus reducing the drift/streaming speed $v_D$ towards $v_A$.

Accompanying wave growth by the CRSI is wave damping. The damping mechanisms vary with
environment, and can be due to ion-neutral collisions \citep{KulsrudPearce69,Soler_etal16}), non-linear
Landau damping \citep{LeeVolk73}, linear Landau damping \citep{FooteKulsrud79,Wiener_etal18},
background  turbulence \citep{FarmerGoldreich04,Lazarian16}, and (potentially) the role of charged
interstellar dust \citep{Squire_etal21}. Strong damping can reduce wave
amplitudes, making them insufficient to isotropize the CRs. Under quasi-linear theory (QLT), it is often
considered that the waves will grow to an extent that partially reduces the effective CR streaming speed
so that wave growth balances wave damping in a (quasi-) steady state
\citep{KulsrudCesarsky71,Skilling71,Wiener_etal13}.

CR feedback is often studied at macroscopic level via CR (magneto-)hydrodynamics. The equations
are obtained by taking moments at the Fokker-Planck equation for CR transport
\citep{Skilling71,Skilling75a,Schlickeiser02} and integrating over the CR momentum (grey approximation).
The equations are expressed for total CR energy
\citep{McKenzieVoelk82,GuoOh08,Pfrommer_etal17}, and more recently also for total CR energy flux
\citep{JiangOh18,ThomasPfrommer19,Chan_etal19,Thomas_etal21}, together with the interaction source
terms between gas and the CRs due to CR scattering by the waves. CR scattering is modeled with a
prescribed transport coefficient (scattering rate or diffusion coefficient), either isotropic
\citep{Uhlig_etal12,Booth_etal13,SalemBryan14,Simpson_etal16,Wiener_etal17} or field aligned
\citep{Hanasz_etal13,Girichidis_etal16,Pakmor_etal16,Ruszkowski_etal17}, and is usually
treated as a constant that is independent of the physical conditions in the system.
Moreover, CRs are often considered as streaming down the CR pressure gradient at a certain streaming
speed that is artificially prescribed \citep[e.g.,][]{Ruszkowski_etal17,Wiener_etal17}. As a result, there are
substantial degrees of freedom in setting the diffusion coefficients and streaming speeds, which can lead to
dramatically different outcomes and hence substantial uncertainties \citep[e.g.][]{Hopkins_etal21b}.

Most uncertainties in these macroscopic studies should be resolved at the microscopic level.
We note that CRs with a given momentum can interact with waves over a wide range of wavelengths
depending on their pitch angle. The interaction of a CR population (with an energy distribution) with waves
thus involves complex convolution of wave-particle interactions. The conventional treatment with QLT
has to make strong simplifying assumptions in deriving the scattering coefficients, leaving major
uncertainties.
A kinetic approach is essential to capture the complex interplay between CR particles and waves, determine the
full wave spectrum, and hence the CR transport coefficients.

Recently, we carried out the first numerical study of the CRSI, with parameters compatible with those typical
in the ISM \citep{Bai_etal19} (hereafter \citetalias{Bai_etal19}). Keys to this work include the use
of magnetohydrodynamic-particle-in-cell (MHD-PIC) method \citep{Bai_etal15}, which substantially alleviates
the issue of scale separation encountered in conventional PIC methods (see \citealp{HolcombSpitkovsky19}),
and the implementation of the $\delta f$ weighting scheme, which dramatically reduces particle noise. We
further incorporated ion-neutral damping in our follow up work \citep{Plotnikov_etal21}.
With a homogeneous and periodic simulation box, we accurately verified the linear growth
rate over a wide range of wavelengths, and precisely tracked the quasi-linear evolution of the particles over
a wide range of momenta. We also found that the main obstacle for isotropizing the CRs lies in the crossing
of 90$^\circ$ pitch angle where certain nonlinearity must be involved to reflect particles.

We note that realistically, a steady, saturated state of the CRSI is achieved by balancing the driving
force of a CR pressure gradient with wave damping.
While the periodic setup is useful for verifying the basic physics of the CRSI, such a steady state can never
be established: the system saturates as the free-energy from CR streaming is exhausted (i.e., CRs are
isotropized in the wave frame).

In this paper, we aim to achieve such a steady state to enable first-principle measurement of CR transport coefficients.
For this purpose, we have designed a novel simulation framework which we term as ``streaming box"
that imposes a fixed CR pressure gradient that constantly drives the growth of the CRSI. It demands
simultaneously resolving the gyro-resonant scale while accommodating multiple CR mean free paths,
making it is truly multi-scale in nature.
As a first study, this paper provides the basic framework, as well as initial results by considering
ion-neutral damping. We anticipate this framework to open up the window for more systematic future
investigations that would eventually provide a comprehensive subgrid model of microscopic CR transport coefficients as a function of local (but macroscopic) environmental parameters. Such a subgrid model, when fed to CR (magneto-)hydrodynamics, is highly desirable for understanding CR transport and feedback on macroscopic scales over a wide range of astrophysical systems.

This paper is organized as follows. We start by describing the basic streaming box
framework and simulation setup in Section \ref{sec:streamingbox}. A theoretical
framework based on QLT is presented in Section \ref{sec:framework} to help
interpret simulation results. In Section \ref{sec:fid}, we show simulation results
mainly from our fiducial runs. This is followed by a survey of results on CR
scattering rates from all simulations in Section \ref{sec:param}.
Our results are further discussed in Section \ref{sec:discussion}, connecting
to CR (magneto-)hydrodynamics. We conclude in Section \ref{sec:conclusion}.
Additional numerical aspects of our simulations are described in the Appendices.

\section[]{Simulation Setup with Streaming Box}\label{sec:streamingbox}

In this section, we describe a novel simulation framework that we term ``streaming box",
which has unique advantages for simulating the CRSI towards saturation that enables
measuring CR scattering rates.
Detailed implementation will be given in Appendix \ref{app:implementation}.

\begin{figure}
    \centering
    \includegraphics[width=85mm]{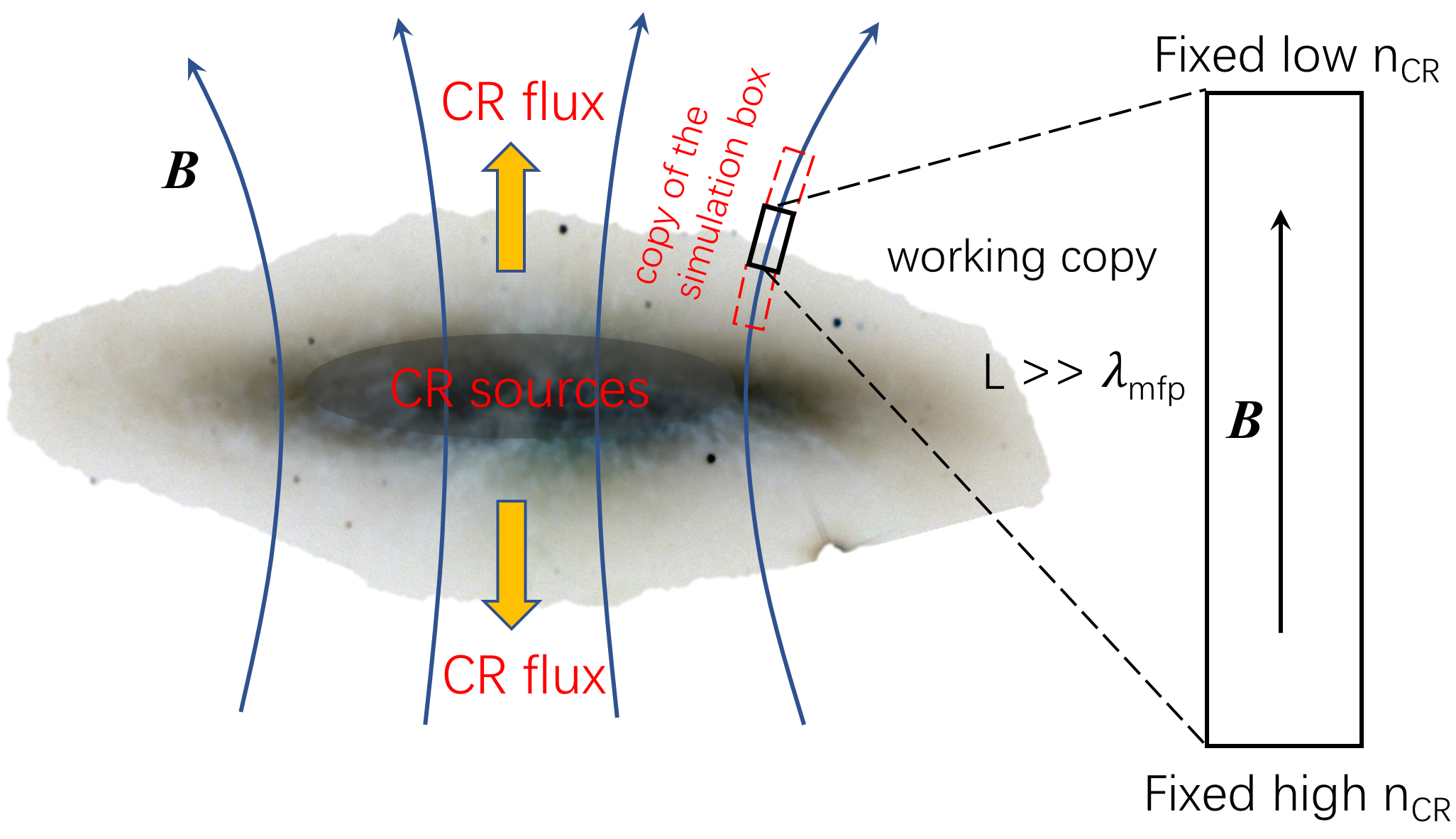}
  \caption{Schematic illustration of the streaming box framework. It can be considered as a local
  patch embedded in a global setting where CRs escape from the source downward the CR
  pressure gradient along magnetic fields on macroscopic scales, e.g., in a star-forming
  galaxy. The box employs periodic boundary conditions for the background gas, while maintaining
  the global CR pressure gradient by imposing fixed but different CR number densities at the
  two boundaries. This creates and drives CR streaming through the box, triggering the CRSI
  to achieve steady state by balancing wave growth and damping. In the
  meantime, the box size should be sufficiently long to accommodate multiple CR mean free paths at
  saturated wave amplitudes. See Section \ref{sec:streamingbox} for details.}\label{fig:schematic}
\end{figure}

\subsection[]{General description of the streaming box framework}\label{ssec:framework}

The streaming box is a Cartesian simulation box which can be considered as a local
patch in a global system where the CRs stream downward a CR pressure gradient (see
Figure \ref{fig:schematic}). The box is aligned with the CR pressure gradient and, in
this work, we further assume the CR pressure gradient is aligned with the background
magnetic field (this can be relaxed when considering multi-dimensional problems). We
refer to this direction as the $\hat{x}$ axis, with box size being $L_x$.

From a macroscopic perspective, the box size $L_x$ should be sufficiently small so that
the fractional variation in CR pressure across the box is small, and the background gas
properties and magnetic fields can be regarded as approximately uniform.\footnote{Arguably,
the gradient in background gas density may exceed the background CR pressure gradient. However,
it is the CR gradient that plays a driving role that must be kept.}
In other words, $L_x\ll L_{\rm macro}$, where $L_{\rm macro}$ is some macroscopic scale
where CR or background gas properties vary significantly.
From microscopic perspective, on the other hand, the box must be long enough to well
accommodate the CR mean free path, or $L_x\gg\lambda_{\rm mfp}$, which is essential for
CRs to be sufficiently scattered to exhibit fluid-like behavior.

Given $L_x\ll L_{\rm macro}$, the initial state of the background gas is set to be uniform, employing
periodic boundary conditions. For CRs, a pressure gradient is enforced by injecting CR particles
from the two box boundaries but at different rates, and is practically achieved by fixing CR
number densities at the boundaries (detailed implementation is described in Appendix
\ref{app:injection}). In other words, the box is essentially connected to neighboring
``boxes" that share similar conditions and maintain similar CR pressure gradient, as depicted in
Figure \ref{fig:schematic}. We denote the CR number densities at the higher and lower sides to be
$n_{\rm CR}^{\rm hi}$ and $n_{\rm CR}^{\rm lo}$, respectively, and further define
\bgeq
n_{\rm CR}^{\rm ctr}=(n_{\rm CR}^{\rm hi}+n_{\rm CR}^{\rm lo})/2\ ,\quad
\Delta n_{\rm CR}=n_{\rm CR}^{\rm hi}-n_{\rm CR}^{\rm lo}\ ,
\edeq
as the mean CR number density and density variation in the box.
As a local approach, we demand $\Delta n_{\rm CR}\ll n_{\rm CR}^{\rm ctr}$. To facilitate
discussion, let our coordinate be chosen such that the left/right boundaries are at $x=0$ and
$x=L_x$, respectively, with high CR pressure on the left ($x=0$) side. Therefore, we anticipate a
steady state CR number density profile $n_{\rm CR,0}(x)$ to be
\bgeq\label{eq:ncrx}
n_{\rm CR,0}(x)=n_{\rm CR}^{\rm hi}-(x/L_x)\Delta n_{\rm CR}\ .
\edeq

A more subtle requirement is that $\Delta n_{\rm CR}/n_{\rm CR}^0\gg v_A/c$. This is
because to trigger the CRSI in the simulation box, there must be some initial anisotropy
that exceeds $\sim v_A/c$. The upper limit of anisotropy that can be achieved in the box
is the free-streaming limit, which would yield an effective streaming speed on the order of
$(\Delta n_{\rm CR}/n_{\rm CR}^0)c$ (for relativistic CRs).
Therefore, we must allow for larger $\Delta n_{\rm CR}/n_{\rm CR}^0$ ratio to accommodate
sufficient range of CR streaming speeds.

To summarize, the basic requirements in the streaming box setting are that
$\lambda_{\rm mfp}\ll L_x\ll L_{\rm macro}$, and $v_A/c\ll\Delta n_{\rm CR}/n_{\rm CR}^0\ll1$.
To some extent, it is analogous to the shearing-box framework in the context of accretion
disk simulations \citep{HGB95}, where the background gradient is ignored to employ
periodic-like boundary conditions, while retaining the driving force (i.e., rotation and shear).
Compared to the conventional periodic box setup in \citetalias{Bai_etal19}, the streaming box
framework is more computationally demanding in requiring
$L_x\gg\lambda_{\rm mfp}\sim(\delta B/B)^{-2}R_g$, where $R_g$ is CR gyroradius, while the
periodic box simply requires $L\gg R_g$ (to properly capture quasi-linear diffusion). We note that
the resolution requirement in a periodic box is already demanding, especially due to the need to
enable reflection across $90^\circ$ pitch angle (see \citetalias{Bai_etal19}), the streaming box
setup thus mandates accommodating a hierarchy of well separated scales:
$\Delta x\ll R_g\ll\lambda_{\rm mfp}\ll L_x$. Therefore, the simulations are most affordable in 1D,
which is our starting point to investigate the rich kinetic physics.

\subsection[]{The MHD-PIC method}\label{ssec:mhdpic}

Our simulations are conducted using the Athena MHD code \citep{Stone_etal08},
supplemented with CR particles to become an MHD-PIC code \citep{Bai_etal15}. 
It solves MHD equations for background gas, and simultaneously integrates CR particle
trajectories in the MHD electromagnetic field, with backreaction on the background
gas. In conservative form, the MHD equations read (\citetalias{Bai_etal19})
\begin{equation}\label{eq:gascont}
\frac{\pa\rho}{\pa t}+\nabla\cdot(\rho\mb{v}_g)=0\ ,
\end{equation}
\begin{equation}
\begin{split}
\frac{\pa(\rho\mb{v}_g)}{\pa t}+\nabla\cdot(&\rho{\mb v}_g^T{\mb v}_g
-{\mb B}^T{\mb B}+P^*{\sf I})\\
=&-(qn_{\rm CR}{\boldsymbol{\mc{E}}}+
{\mb J}_{\rm CR}\times{\mb B})\ ,\\
\end{split}
\label{eq:gasmotion}
\end{equation}
\bgeq
\frac{\pa{\mb B}}{\pa t}=\nabla\times({\mb v}_g\times{\mb B})\ ,\label{eq:ind}
\edeq
where $P^*\equiv P_g+B^2/2$, ${\mc{E}}\equiv-{\mb v}_g\times{\mb B}$ is the
electric field, ${\sf I}$ is the identity tensor.
In the above, $\rho$, ${\mb v}_g$, $P_g$ are gas density, velocity, and pressure. 
Note we have adopted the units where magnetic permeability is unity so that factors
of $(4\pi)^{-1/2}$ that would otherwise appear with magnetic fields are eliminated.
As an initial study, we adopt an isothermal equation of state $P=\rho c_s^2$, with
isothermal sound speed $c_s$, avoiding additional
complications from heating and cooling (see Appendix \ref{app:heating} for further discussion).
We use the Roe Riemann solver
\citep{Roe81} and third-order reconstruction in characteristic variables.

For CR particles, let $m$ and $q$ be particle mass and charge. Given particle
$j$, let its velocity and momentum be ${\mb V}_j$ and ${\mb p}_j$. For notational
convenience, we omit mass in the definition of particle momentum, so
that they are related by ${\mb p}_j=\gamma_j{\mb V}_j$, with Lorentz factor
\begin{equation}\label{eq:Lorentz}
\gamma(p_j)=\frac{\sqrt{{\mathbb C}^2+p_j^2}}{\mathbb C}
=\frac{\mathbb C}{\sqrt{{\mathbb C}^2-V_j^2}}\ ,
\end{equation}
where ${\mathbb C}$ is the (numerical) speed of light for CR particles
(valid as long as ${\mathbb C}\gg v_A$ and $c_s$)\footnote{While one can
rescale the MHD-PIC equations to conform to the reduce-speed-of-light
approximation \citep{JiHopkins22}, we emphasize that the present
formulation is fully valid as long as it ensures sufficient scale separation,
and we will see that the dimensionless scattering coefficients are
independent of ${\mathbb C}$.},
and particle energy is $E(p)\equiv\gamma(p){\mathbb C}^2$.
The particle equation of motion is given by
\begin{equation}
\frac{d\mb{p}_j}{d t}
=\bigg(\frac{q}{m}\bigg)\bigg({\boldsymbol{\mc{E}}}
+{\mb V}_j\times{\mb B}\bigg)\ ,\label{eq:CRmotion}
\end{equation}
and is solved by the standard Boris integrator \citep{Boris70}, together with a standard
triangular-shaped cloud (TSC) scheme \citep{BirdsallLangdon05} for interpolation.
We note that in all occasions,
$q$ and $m$ can be combined to a single factor
$(q/m)$ characterizing the charge-to-mass ratio (\citealp{Bai_etal15}, \citetalias{Bai_etal19}).
In particular, the CR (non-relativistic) cyclotron frequency is given by
$\Omega_c=(q/m)B$.

Backreaction from CRs on the gas is reflected in Equation (\ref{eq:gasmotion}) through
the Lorentz force experienced from the CR population. Defining $f(t, {\mb x}, {\mb p})$ as
the CR momentum distribution function (DF), the CR number density $n_{\rm CR}$ and
current density ${\mb J}_{\rm CR}$ are given by
$n_{\rm CR}(t, {\mb x})=\int f(t, {\mb x}, {\mb p})d^3{\mb p}$ and
${\mb J}_{\rm CR}(t, {\mb x})=q\int {\mb v}f(t, {\mb x}, {\mb p})d^3{\mb p}$.
Assuming background gas has the same composition as CR particles, the magnitude of CR
feedback is determined by the normalization of the DF, characterized by
$n_{\rm CR}/n_i$, where $n_i=\rho/m$ is the number density of the background gas.

\subsection[]{Simulation setup}\label{ssec:setup}

Our streaming box simulations follows a 1D setup along the $\hat{x}$ direction, with uniform
background gas density $\rho_0=1$, background magnetic field ${\mb B}_0=B_0\hat{x}=1$.
This gives a background Alfv\'en speed $v_A=B_0/\sqrt{\rho_0}=1$. We carry our simulations
in the frame of forward propagating Alfv\'en waves, and hence the background gas velocity is set
to ${\mb v}_{g0}=-v_A\hat{x}$ (see Appendix \ref{app:frame} for a discussion on the choice
of simulation frames). We choose the isothermal sound speed to be $c_s=1$, so
that ion thermal pressure and magnetic pressures are about in equipartition.
On top of the background, we initialize a spectrum of Alfv\'en waves as seed
perturbations, with $I(k)$ being the wave intensity at wave number $k$, normalized by
\begin{equation}\label{eq:wavespec}
E_{\rm wave}\equiv\int I(k)dk=\frac{1}{2}\frac{\langle\rho\delta v^2+\delta B^2\rangle}{\rho_0v_{A0}^2}
\approx\frac{\delta B^2}{B_0^2}\ ,
\end{equation}
where angle bracket represents spatial average, and $E_{\rm wave}$ denotes
dimensionless wave energy density.
We initialize the simulations with four Alfven wave modes (left/right handed with
forward/backward propagation) over all wavelengths that can fit into the simulation box,
whose wave intensities are fixed at $kI(k)=10^{-6}$ for all modes with random phases.
Details of wave decomposition are described in Appendix A of \citetalias{Bai_etal19}. 

For CR particles, we set their charge-to-mass ratio $q/m=1$ so that their cyclotron frequency
$\Omega_c=qB_0/m=1$. This choice determines the length unit in the simulations to be
$d_i = v_A/\Omega_c=1$, which we note corresponds to the ion inertial length $c/\omega_{pi}$ for the
background plasma.\footnote{This correspondence assumes the CRs and background plasmas share
the same composition. Note that we use $d_i$ as our length units mainly for convenience, which is more
familiar to the plasma physics community, while we emphasize that our simulations do not capture the
kinetic physics of the background plasmas.}
Our MHD-PIC framework is generally applicable on scales greater than $d_i$.
We choose the numerical speed of light to be $\mathbb{C}=300v_{A}=300$, the same as in
\citetalias{Bai_etal19} to achieve sufficient scale separation ($\mathbb{C}\gg v_A$). 

We take the mean/central background CR DF $f_0^{\rm ctr}(p)$ to be a
$\kappa$ distribution, given by
\begin{equation}
f_0^{\rm ctr}(p)=\frac{n_{\rm CR}^{\rm ctr}}{(\pi\kappa p_0^2)^{3/2}}
\frac{\Gamma(\kappa+1)}{\Gamma(\kappa-\frac{1}{2})}
\bigg[1+\frac{1}{\kappa}\bigg(\frac{p}{p_0}\bigg)^2\bigg]^{-(\kappa+1)}\ ,\label{eq:kappadist}
\end{equation}
and in our streaming box setting, the full distribution function has a
spatial dependence, given by
\bgeq
f_0(x,p)=\frac{n_{\rm CR, 0}(x)}{n_{\rm CR}^{\rm ctr}}f_0^{\rm ctr}(p)\ .\label{eq:f0x}
\edeq
This DF is mainly characterized by $p_0$ and $\kappa$. It
approaches a constant at $p\ll p_0$ and approaches a power-law distribution
$f\propto p^{-2(\kappa+1)}$ a $p\gg p_0$.
Just as in \citetalias{Bai_etal19}, we choose $\kappa=1.25$, corresponding to
$f_0^{\rm ctr}(p)\propto p^{-4.5}$ at large $p$, and set $p_0=\mathbb{C}=300$,
so that the DF peaks at trans-relativistic CR energies. The resonant wavelength for the
bulk CR population is then
$\lambda_0=2\pi p_0/\Omega_c=1.885\times10^3d_i$,
achieving good scale separation from $d_i$.
In our simulations, we use the $\delta f$ method which dramatically reduces noise and
improves the accuracy, as implemented in \citetalias{Bai_etal19} for a homogeneous
background $f_0(p)$. We discuss in Appendix \ref{app:deltaf} about its generalization to
the streaming box framework with the background DF inhomogeneous.

To inject CRs following the DF (\ref{eq:kappadist}), (\ref{eq:f0x}), similar to \citetalias{Bai_etal19},
we divide the momentum space into 16 bins from $0.01p_0$ to $100p_0$, with four bins per dex.
An equal number ($N_p$) of CR particles per cell per bin are injected which guarantees a sufficient
number of particles in each bin.
Within each bin in each cell, particles are sampled according to $f_0(x,p)$ in that
momentum range (given by (\ref{eq:kappadist}) and (\ref{eq:f0x})).
By default, we set $N_{\rm ppc}=96$ particles per cell (or $N_{\rm ppcb}=6$ particles per cell per bin)
for a particle DF at the center of the box $f_0^{\rm ctr}$,
and adjust the number of particles per cell based on the cumulative CR column density from one of
the boundaries to accommodate a CR number density gradient. CR injection at the two boundaries
is described in Appendix \ref{app:injection}.

Among several wave damping mechanisms (mentioned in Section \ref{sec:intro}),
we consider ion-neutral damping that dominates in weakly ionized gas as a first study.
In this case, the background gas that we simulate represents the ion-fluid. The neutral fluid is
considered to be inactive, staying at constant background velocity ${\mb v}_{g0}$ (same as the
initial ion velocity).
We have studied the CRSI with ion-neutral damping in a periodic box in \citet{Plotnikov_etal21}.
The damping term is implemented by operator-splitting as
\bgeq
\frac{\pa{\mb v}_g}{\pa t}=-\nu_{\rm in}({\mb v}_g-{\mb v}_{g0})\ ,
\edeq
where the ions collide with neutrals at frequency
$\nu_{\rm in}=n_n\langle\sigma v\rangle[m_n/(m_n+m_i)]$, here taken to be constant, which
drives the ion velocity towards ${\mb v}_{g0}$. 
In the above, $n_n$ is the number density of the neutrals, and $m_i$, $m_n$ are the
mass of individual ions and neutrals, and $\langle\sigma v\rangle$ is the rate of momentum
transfer. This formulation is valid in the high-frequency regime where
$\omega=kv_A\gg\nu_{\rm in}$, generally applicable in the ISM.

\subsection[]{Main parameters and list of runs}

\begin{table*}
\caption{List of main simulation runs}\label{tab:params}
\begin{center}
\begin{tabular}{c|ccc|cc|c|cccc}\hline\hline
 Run & $\nu_{\rm in}$ &  Box size & $\Delta n_{\rm CR}/n_{\rm CR}^{\rm ctr}$ &Resolution & $N_{\rm ppc}$ & Runtime  & $E_{\rm wave}$ & $\Delta v_g$ & $L_{\rm CR}$ & $\alpha_{\rm sca}$ \\
   &     ($10^{-4}\Omega_c$) & $(L_x/d_i)$ & & ($\Delta x/d_i$) & & ($\Omega_c^{-1})$ & ($\rho_0v_A^2$) & ($v_A$) & ($10^7d_i$) & \\\hline
Fid & 1.0 &  $10^7$ & 0.4 & $5$ & $96 $ & $2.0\times10^5$ & $2.70\times10^{-3}$ & $2.77\times10^{-3}$ & 2.39 & 0.0544 \\\hline
Lores & 1.0 & $10^7$ & 0.4 & $10$ & $128 $ & $2.5\times10^5$ & $2.64\times10^{-3}$ & $2.71\times10^{-3}$ & 2.38 & 0.0529 \\
Hires & 1.0 & $10^7$ & 0.4 & $2.5$ & $64 $ & $1.6\times10^5$ & $2.67\times10^{-3}$ & $2.70\times10^{-3}$ & 2.40 & 0.0554 \\\hline
Lograd & 1.0 & $10^7$ & 0.24 & $5$ & $96 $ & $2.5\times10^5$ & $1.27\times10^{-3}$ & $1.29\times10^{-3}$ & 4.92 & 0.0534 \\
Higrad & 1.0 & $10^7$ & 0.56 & $5$ & $96 $ & $1.8\times10^5$ & $3.98\times10^{-3}$ & $4.07\times10^{-3}$ & 1.63 & 0.0557 \\\hline
Short & 1.0 & $5\times10^6$ & 0.4 & $5$ & $96 $ & $1.8\times10^5$ & $5.41\times10^{-3}$ & $5.68\times10^{-3}$ & 1.21 & 0.0531 \\\hline
Nu05 & 0.5 & $10^7$ & 0.4 & 5 & $96$ & $1.8\times10^5$ & $5.53\times10^{-3}$ & $5.88\times10^{-3}$ & 2.31 & 0.0505 \\
\hline\hline
\end{tabular}
\end{center}
Fixed parameters: ${\mathbb C}=300v_{A0}$, $p_0=300v_{A0}$, 
$n_{\rm CR}/n_i=10^{-4}$.
Values in the last four columns correspond to wave intensity, ion-neutral drift velocity, CR scale length
and dimensionless CR scattering coefficient, measured at final state of the simulations. Note that the latter
two quantities are measured for CRs in the momentum range between $0.1p_0$ and $10p_0$.
\end{table*}

Unless otherwise noted, we also fix most of the main control parameters to our simulations to be
\begin{itemize}\itemsep1pt
\item[$-$] CR number density ratio $n_{\rm CR}^{\rm ctr}/n_i=10^{-4}$.
\item[$-$] CR variation $\Delta n_{\rm CR}/n_{\rm CR}^{\rm ctr}=0.4$.
\item[$-$] Simulation box size $L_x=10^7d_i$.
\item[$-$] Ion-neutral damping rate $\nu_{\rm in}=10^{-4}\Omega_c$.
\end{itemize}
Our choice of box size corresponds to $L_x\sim5300\lambda_{\rm res}$, and hence we
can accommodate particles with mean free path up to
$\lambda_{\rm mfp}\sim10^3\lambda_{\rm res}$. This allows us to achieve relatively small wave
amplitudes approaching realistic conditions of $\delta B/B_0\ll1$.
With these, we define the CR scale length
\bgeq\label{eq:LCR}
L_{\rm CR}\equiv \frac{n_{\rm CR}}{{\mb b}\cdot\nabla n_{\rm CR}}
\approx\frac{n_{\rm CR}^{\rm ctr}}{\Delta n_{\rm CR}}L_x\ ,
\edeq
which characterizes the spatial gradient of the CR distribution (${\mb b}$ is the unit vector along the
magnetic field).
Generally, we anticipate $L_{\rm CR}\approx (n_{\rm CR}^{\rm ctr}/\Delta n_{\rm CR})L_x$,
while in practice, as discussed in Appendix \ref{app:frame} and shown in Section \ref{ssec:evolve},
the final CR density/pressure gradient will show minor deviations.
We thus only conduct measurements around the simulation box center, and report the results
based on the locally measured values of $L_{\rm CR}$ there.

The list of all our simulations runs are given in Table \ref{tab:params}.
By default, we choose the cell size to be $\Delta x=5d_i$,
but also conduct one higher-resolution run with $\Delta x=2.5d_i$ (run name ``Hires") and one
lower-resolution run with $\Delta x=10d_i$ (run name ``Lores"). This choice corresponds to
$\lambda_{\rm res}/\Delta x\sim187-750$.
We will demonstrate that this resolution is sufficient to allow particles to overcome the $\mu=0$
barrier,
and hence we are able to capture the full
scattering processes to properly measure the resulting scattering rates.
Our simulations are typically run for $1.6-2.5\times10^{5}\Omega_c^{-1}$, which is about $4-7$
box-crossing time for typical particles, generally sufficient to achieve a steady state.

\section[]{Theoretical Framework}\label{sec:framework}

Before showing simulation results,
we lay down a theoretical framework for measuring CR transport coefficients
(scattering rates) from our simulations, to be compared with those obtained from QLT.

\subsection[]{Linear instability of waves}\label{ssec:linear}

Linear growth of CRSI is usually derived in a homogeneous setting with constant CR streaming
speed $v_D$, namely, the CR DF is isotropic in the drift frame. 
Here, we only summarize the necessary results needed for setting up and analyzing
the simulations. For this purpose, we define the wave-frame streaming speed $v_s$ as $v_s=v_D-v_A$, Further details can be found in \citetalias{Bai_etal19} and \citet{Plotnikov_etal21}.
The CRSI drives growth of forward-propagating, both left and right polarized Alfv\'en waves. They
share the same growth rate given by
\begin{equation}\label{eq:growth}
\Gamma_{\rm CR}(k)\approx\frac{\pi}{4}\frac{n_{\rm CR}}{n_i}\Omega_c
\frac{v_s}{v_A}Q(k)\ ,
\end{equation}
where\footnote{Our definition of $Q(k)$ is a factor $2/\pi$ of $Q_2(k)$ in \citetalias{Bai_etal19}.}
\begin{equation}\label{eq:Q2}
Q(k)\equiv
\int_{p_{\rm res}}^{\infty}dp\frac{4\pi p^2f_0(p)}{n_{\rm CR}}
\bigg(\frac{p_{\rm res}}{p}\bigg)\ ,
\end{equation}
and $p_{\rm res}(k)=\Omega_c/k$ is the minimum CR momentum to resonate with an
Alfv\'en wave with wave number $k$ (at zero pitch angle). Here we use $n_{\rm CR}$ and
$f_0$ to ease our notation, while they should be understood as $n_{\rm CR}^{\rm ctr}$ and
$f_0^{\rm ctr}$ in our setting. With the $\kappa$ distribution (\ref{eq:kappadist}),
we have
\begin{equation}
Q(k)=\frac{2}{\kappa^{3/2}\sqrt{\pi}}\frac{\Gamma(\kappa+1)}{\Gamma(\kappa-\frac{1}{2})}
\frac{1}{s(k)[1+1/(\kappa s(k)^2)]^\kappa}\ ,\label{eq:Q2kappa}
\end{equation}
where
$s(k)\equiv p_0/p_{\rm res}=kp_0/\Omega_c$. The growth rate peaks at $s=1$ or
$k_0\equiv\Omega_c/p_0$,
namely, at the resonant wavelength of particles with momentum $p_0$, where $Q$ is slightly
smaller than order unity ($\sim0.36$ for $\kappa=1.25$). Asymptotically, the growth rate scales as
$\Gamma(k)\sim k^{2\kappa-1}$ at long-wavelength limit, and as $\Gamma(k)\sim k^{-1}$ at
short-wavelength limit (see Figure 3 of \citetalias{Bai_etal19}).

With ion-neutral damping, and focusing on the high-frequency limit
($\omega\gg\nu_{\rm in}$), the growth rate becomes
\bgeq\label{eq:gam_tot}
\Gamma_{\rm tot}(k)\approx\Gamma_{\rm CR}(k)-\frac{\nu_{\rm in}}{2}\ .
\edeq
Therefore, all waves are simultaneously damped at a constant rate of $\nu_{\rm in}/2$.
No wave growth is expected for 
$\nu_{\rm in}>2\Gamma_{\rm CR}(k_0)$. Examples can be found in Figure 5 of
\citet{Plotnikov_etal21}.

\subsection[]{Fokker-Planck equation in the wave frame}\label{ssec:FP}

As the CRSI drives wave growth, CRs undergo pitch angle diffusion in the Alfv\'en wave
frame. The DF can be expressed as a function of $(p,\mu)$, where $\mu\equiv\cos\theta$
for pitch angle $\theta$. Because the electric field is zero in the wave frame, 
results can be analyzed on a momentum-by-momentum basis. 
Starting from the Fokker-Planck equation
(e.g., \citealp{Jokipii66,KulsrudPearce69})
\begin{equation}\label{eq:qld}
\frac{\pa f}{\pa t}+\mu V\frac{\pa f}{\pa x}=
\frac{\pa}{\pa\mu}\bigg(D_{\mu\mu}\frac{\pa f}{\pa\mu}\bigg)\ ,
\end{equation}
where $D_{\mu\mu}$ the pitch angle Fokker-Planck coefficient, given by
\begin{equation}\label{eq:Dmumu}
D_{\mu\mu}(p,\mu)\equiv\bigg\langle\frac{\Delta\mu^2}{2t}\bigg\rangle=\frac{1-\mu^2}{2}\nu_{\mu}(p,\mu)\ ,
\end{equation}
where $\Delta\mu$ denotes $\mu(t_0+t)-\mu(t_0)$ for individual particles at two different
times separated by $t$, followed by an ensemble average denoted by $\langle\cdot\rangle$,
and $\nu_{\mu}(p,\mu)$ is the pitch angle scattering rate.

In QLT, the pitch angle scattering rate is given by
\begin{equation}\label{eq:nu_qld}
\nu_{\mu,\rm QL}(p,\mu)=\pi\Omega k_{\rm res}I^{L/R}(k_{\rm res})\ ,
\end{equation}
where
$k_{\rm res}\equiv \Omega_c/(p|\mu|)$
is the wave number of the resonant waves. The wave intensity $I(k)$ refers to
forward-propagating left/right polarized Alfv\'en waves, with left modes $I^L$ chosen for
$\mu>0$ and right modes $I^R$ chosen for $\mu<0$. Also note that
$\Omega=\Omega_c/\gamma$ is the gyro-frequency.

A very useful result can be obtained by assuming steady state.
Following a more formal derivation in Appendix \ref{app:derivenu}, we obtain
\bgeq\label{eq:nu_ss}
\nu_\mu(p,\mu)\frac{\pa f}{\pa\mu}\approx-V(p)\frac{\pa f}{\pa x}\ ,
\edeq
This relation will allow us to directly measure $\nu_\mu(p,\mu)$ at the saturated state of our
simulations. By comparing this result with (\ref{eq:nu_qld}), it will allow to further constrain
the contributions from non-linear effects.

\subsection[]{Moment equations}

By integrating Equation (\ref{eq:qld}) over $\mu$, we obtain
\bgeq\label{eq:df0}
\frac{\pa F_0}{\pa t}+V(p)\frac{\pa F_1}{\pa x}=0\ ,
\edeq
where the zeros and first moments of the DF are
\bgeq
\begin{split}
F_0(x,p)&\equiv\frac{1}{2}\int_{-1}^1 f(x,p,\mu)d\mu\ ,\\
F_1(x,p)&\equiv\frac{1}{2}\int_{-1}^1 \mu f(x,p,\mu)d\mu\ .
\end{split}
\edeq
Note $F_0$ is expected to be close but not necessarily the same as the
$f_0$ part of the $\delta f$ weighting scheme (due to relaxation near the boundaries),
and the first moment $F_1$ reflects the level of anisotropy in the streaming CRs.

By further multiplying $\mu$ to Equation (\ref{eq:qld}) and then integrating over $\mu$,
we obtain
\bgeq\label{eq:df1}
\begin{split}
\frac{\pa F_1}{\pa t}+\frac{V(p)}{3}\frac{\pa F_0}{\pa x}
&=-\int_{-1}^{1}\frac{\pa f}{\pa\mu}\frac{1-\mu^2}{4}\nu_\mu(p,\mu)d\mu\ ,
\end{split}
\edeq
where we have assumed the second moment satisfies
\bgeq
F_2(x,p)\equiv\frac{1}{2}\int_{-1}^1 \mu^2 f(x,p,\mu)d\mu\approx\frac{1}{3}F_0\ ,
\edeq
an assumption that holds when $f$ is close to isotropic, in line with the streaming
box setup.
Still, this equation is not closed as the right hand side of the equation depends
on the details of the DF.

To proceed, the most common
approach is to truncate the expansion in $f(x,p,\mu)$ to lowest order, and write
\bgeq
f(x,p,\mu)\approx F_0(x,p)+3\mu F_1(x,p)\ ,\label{eq:edddef}
\edeq
which is known as the Eddington approximation (e.g. \citealp{ThomasPfrommer19}), and
Equation (\ref{eq:df1}) becomes
\bgeq\label{eq:eddasump}
\begin{split}
\frac{\pa F_1}{\pa t}+&\frac{V(p)}{3}\frac{\pa F_0}{\pa x}\approx-\nu_{\rm Edd}(p)F_1\ ,\\
\nu_{\rm Edd}(p)&\equiv\frac{3}{4}\int_{-1}^1(1-\mu^2)\nu_\mu(p,\mu)d\mu\ .
\end{split}
\edeq
In QLT, if we assume $\nu_\mu(p,\mu)=\nu_\mu(p,-\mu)$, i.e., left/right handed waves
have equal power, and that $I(k)\propto k^{-q}$ over some range of $k$ around $k_0$,
then for a given $p$ around $p_0$, we may write
$\nu_\mu(p,\mu)=\nu_{0}(p|\mu|/p_0)^{q-1}/\gamma(p)$,
where
$\nu_0\equiv\pi\Omega_ck_0I(k_0)$,
and this yields
\bgeq\label{eq:nu2}
\nu_{\rm Edd}(p)\approx\frac{3}{q(q+2)}\frac{\nu_0}{\gamma(p)}\bigg(\frac{p}{p_0}\bigg)^{q-1}\ .
\edeq
In reality, the Eddington approximation does not necessarily hold, as we will show
from our simulations.

In saturated (steady) state of our simulations, we can measure effective scattering rate
$\nu_{\rm sca}(p)$, and the associated CR mean free path $\lambda_{\rm mfp}(p)$, as
\bgeq\label{eq:nueff}
\nu_{\rm sca}(p)\equiv-\frac{V(p)}{3F_1(p)}\frac{\pa F_0(p)}{\pa x}\ \Rightarrow\ 
\lambda_{\rm mfp}(p)=\frac{V(p)}{\nu_{\rm sca}(p)}\ .
\edeq
By comparing $\nu_{\rm sca}(p)$ with $\nu_{\rm Edd}(p)$, we will be able to
assess how deviations from the Eddington approximation affect the estimation of scattering rates.
The parallel diffusion coefficient $D_x(p)$ is related to $\nu_{\rm sca}(p)$ by
\begin{equation}
D_x(p)=\frac{1}{3}\lambda_{\rm mfp}(p)V(p)=\frac{V(p)^2}{3\nu_{\rm sca}(p)}\ .
\end{equation}
Given the one-to-one correspondence with $D_x(p)$, we here primarily use
$\nu_{\rm sca}(p)$ in our discussion.

\subsection[]{CR (magneto-)hydrodynamics in the wave frame}

The moment equations above can be equivalently expressed in terms of
CR hydrodynamics, given by
\bgeq\label{eq:crcont}
\frac{\pa\epsilon_{\rm CR}}{\pa t}+\frac{\pa {\cal F}_{\rm CR}}{\pa x}=0\ .
\edeq
\bgeq\label{eq:fluid1}
\begin{split}
\frac{\pa {\cal F}_{\rm CR}}{\pa t}&+{\mathbb C}^2\frac{\pa P_{\rm CR}}{\pa x}\equiv-\nu_{\rm sca}(p){\cal F}_{\rm CR}\\
=&-4\pi p^3v(p)E(p)\int_{-1}^1 \frac{\pa f}{\pa\mu}\frac{1-\mu^2}{4}\nu(p,\mu)d\mu\ .
\end{split}
\edeq
where the CR energy density $\epsilon_{\rm CR}$,
CR energy flux ${\cal F}_{\rm CR}$, and CR pressure $P_{\rm CR}$, are defined as\footnote{In CR hydrodynamics,
$\epsilon_{\rm CR}$, ${\cal F}_{\rm CR}$ are often defined in observer's frame.
By noting that $(\epsilon_{\rm CR}, {\cal F}_{\rm CR})$ is a four-vector, our
equations can be easily transformed to different frames.}

\bgeq
\begin{split}\label{eq:fluiddef}
\epsilon_{\rm CR}&=\int d^3{\mb p}E(p)f(x,{\mb p})\equiv\int \epsilon_{\rm CR}(p)d\ln p\ ,\\
{\cal F}_{\rm CR}&=\int d^3{\mb p}\mu V(p)E(p)f(x,{\mb p})\equiv\int{\cal F}_{\rm CR}(p)d\ln p\ ,\\
P_{\rm CR}&=\int d^3{\mb p}\mu^2 V(p)pf(x,{\mb p})\equiv\int P_{\rm CR}(p)d\ln p\ ,
\end{split}
\edeq
where
\bgeq\label{eq:efp}
\begin{split}
\epsilon_{\rm CR}(p)&
=4\pi p^3E(p)F_0(p)\ ,\\
{\cal F}_{\rm CR}(p)&
=4\pi p^3 V(p)E(p)F_1(p)\ ,\\
P_{\rm CR}(x,p)&
=\frac{4\pi}{3}p^4V(p)F_0(p)=\frac{1}{3}\frac{V^2}{{\mathbb C}^2}\epsilon_{\rm CR}(p)\ .
\end{split}
\edeq
Note that the definition of CR pressure implicitly assumes that CR anisotropy is small
(${\cal F}_{\rm CR}\ll\epsilon_{\rm CR}\mathbb{C}$, similar to the relation between $F_2$
and $F_0$). Below, without loss of clarity, we do not distinguish between
momentum-integrated $\epsilon_{\rm CR}$ and $\epsilon_{\rm CR}(p)$, etc.

Under CR hydrodynamics, it allows us to define the momentum-integrated effective
scattering rate with a physically-meaningful weighting
\bgeq\label{eq:nufluid}
\nu_{\rm sca}
=\frac{\int\nu_{\rm sca}(p){\cal F}_{\rm CR}d\ln p}{\int {\cal F}_{\rm CR}d\ln p}\\
=-\frac{\frac{\pa}{\pa x}\int V(p)^2\epsilon_{\rm CR}d\ln p}{3\int {\cal F}_{\rm CR}d\ln p}\ .
\edeq

Due to limited dynamical range, our simulations can only provide reliable measurements
of $\nu_{\rm sca}(p)$ over finite range of CR momentum. Therefore, we typically only
conduct the $p-$integral over the range $0.1p_0\leq p\leq 10p_0$. Since CRs over this range
are expected to be more efficiently scattered, it overestimates $\nu_{\rm sca}$ by a factor
of $\gtrsim2$, as will be discussed in Section \ref{ssec:nutot}.

\subsection[]{Piecing together: theoretical expectations}\label{ssec:theoryall}

The saturated state in our streaming box setting is achieved by balancing the rate
of wave growth and damping. 
Such an approach is broadly referred to as being based on QLT in the literature.\footnote{As we only
consider ion-neutral damping here, which is a linear damping mechanism, QLT is in fact barely
utilized in the derivation, making the derivation less uncertain. When non-linear damping mechanisms,
such as non-linear Landau damping is considered, one would have to explicitly account for
amplitude-dependent scattering rates based on QLT. This is left for our future work.}
In practice, however, there are substantial uncertainties pertaining to the exact approximations made to
the derivation, leading to wildly different results. Here we roughly divide the treatments in the literature into
two categories, namely, treating the CRs as a single fluid (Section \ref{sssec:nu_single}) and treating the
CRs momentum by momentum (Section \ref{sssec:nu_mom}).

The starting point is by defining the effective streaming speed $v_s^{\rm eff}$, which is needed to
estimate wave growth rate. From a fluid point of view, 
we can associate $v_s^{\rm eff}$ with the energy flux ${\cal F}_{\rm CR}$ by
\begin{equation}
\begin{split}\label{eq:vseff}
{\cal F}_{\rm CR}(p) &= v_s^{\rm eff}(p)[\epsilon_{\rm CR}(p)+P_{\rm CR}(p)]\ ,\\
{\cal F}_{\rm CR} &= v_s^{\rm eff}(\epsilon_{\rm CR}+P_{\rm CR})\ .
\end{split}
\end{equation}
From the definitions (\ref{eq:efp}), the above can be expressed as 
\bgeq\label{eq:vseff1}
\begin{split}
v_s^{\rm eff}(p)&
=V(p)\frac{3{\mathbb C}^2}{3{\mathbb C}^2+V(p)^2}\frac{F_1(p)}{F_0(p)}\ ,\\
v_{s}^{\rm eff}&
=\frac{\int V(p)p^2E(p)F_1(p)dp}{\int p^2[E(p)+Vp/3]F_0(p)dp}\ .
\end{split}
\edeq
We remind the readers that here $v_s^{\rm eff}$ is defined in the wave frame.

We shall see in the discussion below that the CR scattering rate should be normalized
by 
\bgeq\label{eq:nu_norm}
\nu_{\rm norm}\equiv\bigg(\frac{\mathbb C}{v_A}\bigg)\bigg(\frac{\mathbb C}{L_{\rm CR}\nu_{\rm in}}\bigg)
\bigg(\frac{n_{\rm CR}^{\rm ctr}}{n_i}\bigg)\Omega_c\ ,
\edeq
which reflects the scaling from QLT. We write 
\bgeq
\nu_{\rm sca}=\alpha_{\rm sca}\nu_{\rm norm}\ ,\quad
\nu_{\rm sca}(p)=\alpha_{\rm sca}(p)\nu_{\rm norm}\ .
\edeq
Calibrating $\alpha_{\rm sca}$ and $\alpha_{\rm sca}(p)$ is one major goal of our simulations.

\subsubsection[]{Treating CRs as a single fluid}\label{sssec:nu_single}

In this first approach, there is a single streaming speed $v_s^{\rm eff}$, and transport coefficients are
derived by balancing growth and damping at the most unstable wavelength.

From (\ref{eq:growth}), with the most unstable $k_0=\Omega_c/p_0$, we may approximately express the peak CRSI growth rate as
\begin{equation}\label{eq:growtheff}
\Gamma_{\rm CR}^{\rm eff}(k_0)\approx\frac{\pi}{4}\frac{n_{\rm CR}}{n_i}\Omega_c\frac{v_s^{\rm eff}}{v_A}Q(k_0)
=0.28\frac{n_{\rm CR}}{n_i}\Omega_c\frac{v_s^{\rm eff}}{v_A}\ .
\end{equation}
Balancing the above with ion-neutral damping rate $\Gamma_{\rm damp}=\nu_{\rm in}/2$,
we obtain for the effective streaming speed (see also \citealp{Plotnikov_etal21}):
\bgeq\label{eq:vsexp}
v_{s,1}^{\rm eff}\approx\frac{2}{\pi Q(k_0)}\frac{\nu_{\rm in}}{\Omega_c}\frac{n_i}{n_{\rm CR}}v_A
\approx1.8\frac{\nu_{\rm in}}{\Omega_c}\frac{n_i}{n_{\rm CR}}v_A\ .
\edeq
This is a theoretical expectation
commonly employed in the literature (e.g., \citealp{Wiener_etal13}). In particular, as a linear
damping mechanism, this expected streaming speed is independent of the imposed CR gradient
\citep{Skilling71,KulsrudCesarsky71}.

Combining Equations (\ref{eq:nufluid}) and (\ref{eq:vseff}), we establish the relation
\bgeq
\begin{split}
\nu_{\rm sca,1}^{\rm pred}&=\frac{
\int V(p)^2\epsilon_{\rm CR}d\ln p}{3v_{s,1}^{\rm eff}L_{\rm CR}\int (\epsilon_{\rm CR}+P_{\rm CR})d\ln p}
\equiv\alpha_{\rm sca,1}^{\rm pred}\nu_{\rm norm}\ ,\\
\end{split}\label{eq:nuscaeff}
\edeq
where the dimensionless prefactor is given by
\bgeq\label{eq:alphasca}
\alpha_{\rm sca,1}^{\rm pred}=\frac{\pi}{2}Q(k_0)\frac{\int V(p)^2\epsilon_{\rm CR}d\ln p}
{\int(3{\mathbb C}^2+V(p)^2)\epsilon_{\rm CR}d\ln p}\ ,
\edeq
which we find to be $0.124$ for our choice of the DF.

\subsubsection[]{Treating CRs momentum by momentum}\label{sssec:nu_mom}

Alternatively, one can treat CRs momentum-by-momentum, each having their
own drift speed $v_s^{\rm eff}(p)$. To approximate wave growth rate at wave
number $k$, we treat particles with $p\geq p_{\rm res}(k)$ as sharing the same
streaming velocity $v_s^{\rm eff}(p_{\rm res})$, and write
\begin{equation}
\Gamma_{\rm CR}^{\rm pred}(k)\approx\frac{\pi}{4}\frac{n_{\rm CR}}{n_i}
\Omega_c\frac{v_s^{\rm eff}(p_{\rm res})}{v_A}Q(k)\ ,\label{eq:gamcr_pred}
\end{equation}
where we use the resonance condition $p_{\rm res}=\Omega_c/k$, or
$k_{\rm res}=\Omega_c/p$, and the $Q(k)$ factor approximately encapsulates
the contribution from all particles to the growth at that wavelength. 
Such approach has been employed in the recent literature to study the role of
self-confinement on CR transport
\citep{AmatoBlasi18,Evoli_etal18}, but we note that except for \citet{Armillotta_etal21}
who adopted almost identical formulation, our formula is more accurate
that those commonly employed\footnote{Our formula is largely identical to those
adopted in the literature except for the the $Q(k)$ factor defined in (\ref{eq:Q2}),
as opposed to the commonly-used $n_{\rm CR}(p>p_{\rm res})/n_{\rm CR}$,
or $4\pi p^3f(p)/n_{\rm CR}$.
}.

Balancing (\ref{eq:gamcr_pred}) with damping rate $\Gamma_{\rm damp}=\nu_{\rm in}/2$,
we obtain the predicted streaming speed as
\begin{equation}\label{eq:vdprd}
v_{s,2}^{\rm eff}(p)=\frac{2}{\pi}\frac{\nu_{\rm in}}{\Omega_c}\frac{n_i}{n_{\rm CR}}Q(k_{\rm res})^{-1}v_A\ .
\end{equation}
With Equation (\ref{eq:fluid1}), taking $\pa/\pa t=0$ for steady state, and
expecting $\pa/\pa x \approx -L_{\rm CR}^{-1}$, we find
\bgeq\label{eq:vseff1}
v_{s}^{\rm eff}(p)\approx\frac{V(p)^2}{3\nu_{\rm sca}(p)}\frac{1}{L_{\rm CR}}
\frac{\epsilon_{\rm CR}(p)}{\epsilon_{\rm CR}(p)+P_{\rm CR}(p)}
\edeq
for individual CR momentum.
Equating the above two expressions, we find the predicted scattering rate
\begin{equation}\label{eq:nuscapred}
\nu_{\rm sca,2}^{\rm pred}(p)=\frac{\pi}{2}Q(k_{\rm res})\frac{V(p)^2}{3{\mathbb C}^2+V(p)^2}\nu_{\rm norm}
\equiv\alpha_{\rm sca,2}^{\rm pred}(p)\nu_{\rm norm}\ .
\end{equation}
Note the close similarity of $\alpha_{\rm sca,2}^{\rm pred}(p)$ with (\ref{eq:nuscaeff}) and
(\ref{eq:alphasca}) since they are derived based on the same principles.

Note that it is largely identical to Equation (17) of \citet{Armillotta_etal21} except that
they took the last factor to be $3/4$ assuming $V(p)={\mathbb C}$.
Here, we assume $L_{\rm CR}$ for particles of all energies are the same, set by
$n_{\rm CR}^{\rm ctr}/\Delta n_{\rm CR}L$ in our case.
Given the scalings in our adopted $\kappa$-distribution, we have
$v_{s,2}^{\rm eff}(p)\propto p^{-1}$, $\nu_{\rm sca,2}^{\rm pred}(p)\sim p^3$ at $p\ll p_0$ and
$v_{s,2}^{\rm eff}(p)\propto p^{2\kappa-1}$,
$\nu_{\rm sca,2}^{\rm pred}(p)\sim p^{1-2\kappa}$ at $p\gg p_0$.
Note that with this estimate, $v_{s,2}^{\rm eff}(p)$ diverges at both large and small $p$.

Integrating the above over momentum, we obtain 
\begin{equation}\label{eq:vscprd}
\nu_{\rm sca,2}^{\rm pred}\approx\frac{1}{3L_{\rm CR}}
\frac{\int V(p)^2\epsilon_{\rm CR}(p)d\ln p}{\int [\epsilon_{\rm CR}(p)+P_{\rm CR}(p)]v_{s,2}^{\rm eff}(p)d\ln p}\ ,
\end{equation}
which then yields a relation similar to Equation (\ref{eq:nuscaeff}). However, the resulting
prefactor $\alpha_{\rm sca,2}^{\rm pred}$ does not converge with the upper momentum bound of the
integral, as the denominator scales as $\int pd\ln p$ which quickly diverges (the results are
extremely insensitive to the lower-bound of the integral). In reality, this issue can be resolved
as scattering by external turbulence will take over beyond some momentum, which can be
treated as an effective truncation at some $p_{\rm max}$. We obtain $\alpha_{\rm sca,2}^{\rm pred}=0.046$
for $p_{\rm max}=10p_0$, and $\alpha_{\rm sca,2}^{\rm pred}=0.0064$ for $p_{\rm max}=100p_0$. In
either case, the value of $\alpha_{\rm sca,2}^{\rm pred}$ is much smaller than our previous single-fluid
estimate, again reflecting dramatic uncertainties.

\section[]{Simulation results: fiducial run}\label{sec:fid}

In this section, we focus on our fiducial run and discuss in detail the physics
of the CRSI driven by an imposed background CR gradient.

\begin{figure}
    \centering
    \includegraphics[width=90mm]{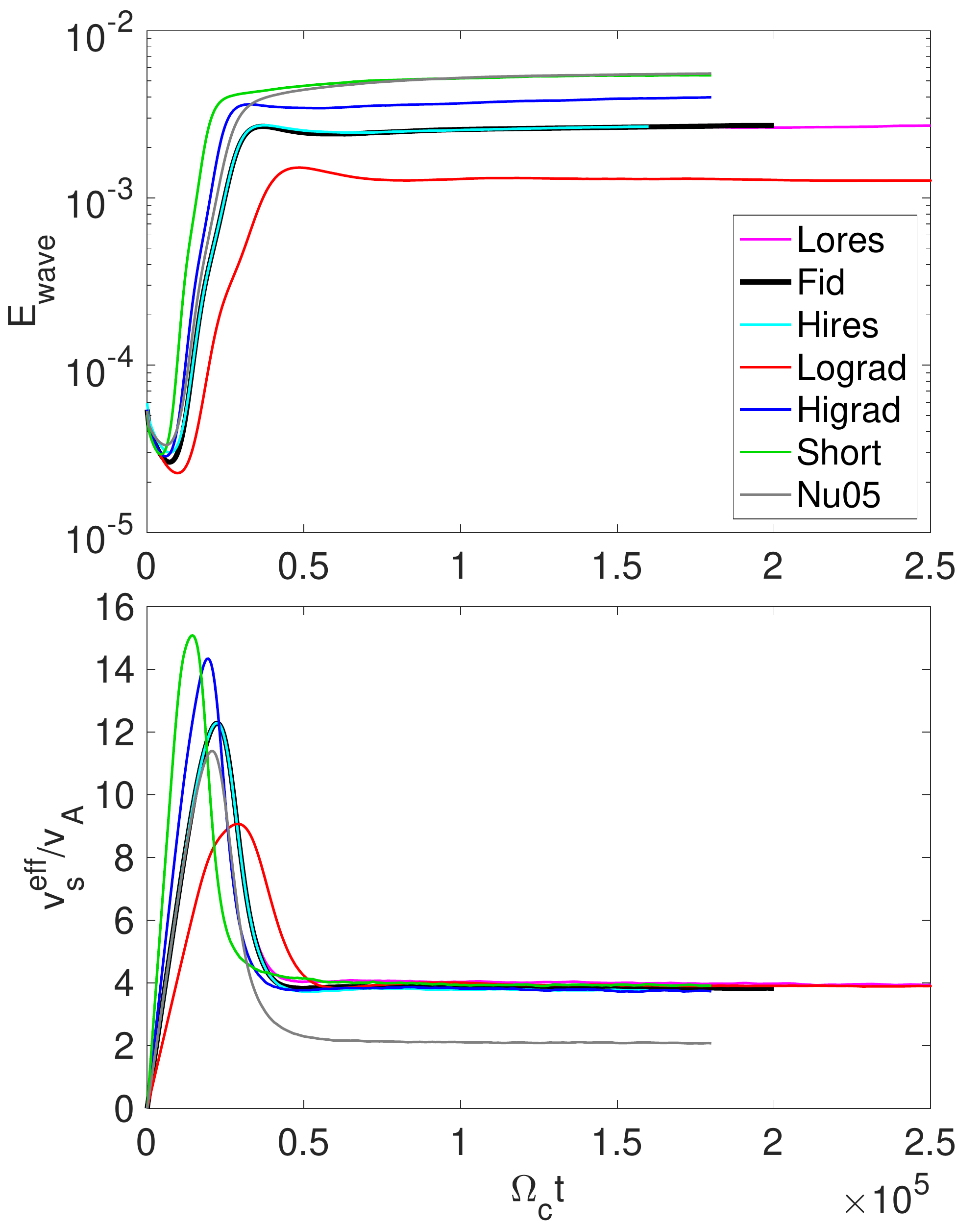}
  \caption{Time evolution of the wave energy density (top) and effective
  streaming speed (bottom) from all our simulations. In the bottom panel,
  we show $v_s^{\rm eff}$ defined by Equation (\ref{eq:vseff}), averaged
  over momentum $0.1p_0<p<10p_0$.}\label{fig:EwvWeff}
\end{figure}

\begin{figure}
    \centering
    \includegraphics[width=90mm]{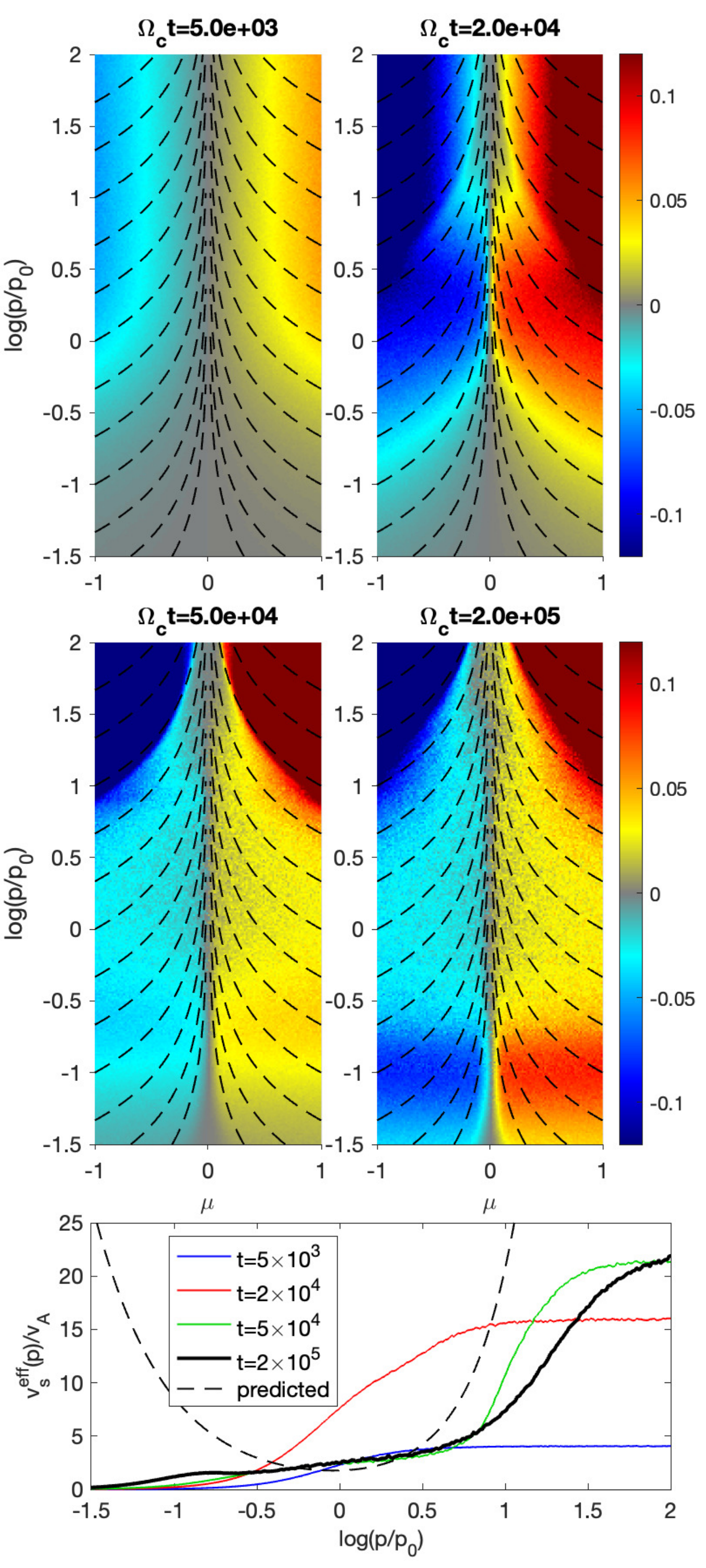}
  \caption{Anisotropy in the CR distribution function, shown as $\delta f(p,\mu)$, at
  four different simulation snapshots from the fiducial run. The bottom panel shows
  the effective streaming speed for each particle momentum at these snapshots,
  over-plotted with the predicted curve given by Equation (\ref{eq:vdprd}) in dashed line.}\label{fig:fpar_hist}
\end{figure}

\subsection[]{Overall evolution}\label{ssec:evolve}

In Figure \ref{fig:EwvWeff}, we show the time evolution of wave energy density
and the effective CR streaming speed $v_s^{\rm eff}$ (measured in the simulation
frame from (\ref{eq:vseff}), i.e., the wave frame) for all simulation runs. Here, the
dimensionless wave energy density is defined as
$E_{\rm wave}=(\rho v_\perp^2+B_\perp^2)/(2\rho v_A^2)$, and we focus on the
the fiducial run in this section (black line).
In the meantime, we show in Figure \ref{fig:fpar_hist} snapshots of the distribution
function measured at the central region of the simulation box in the wave frame. It
is shown in the form of $\delta f/f_0$ as a function of $(p,\mu)$, where grey indicates
isotropy.

Initially, wave energy decreases because of wave damping. Particles are initially
isotropic in the wave frame with $v_s^{\rm eff}=0$. The background pressure gradient 
leads to rapid development of CR streaming, with $v_s^{\rm eff}$ increasing
linearly in time. This is better viewed from the first panel of Figure \ref{fig:fpar_hist},
showing an excess of particles with $\mu>0$ and deficit of particles with $\mu<0$.
This triggers the CRSI, and wave energy undergoes exponential growth shortly afterwards. 

As wave amplitude increases, wave-particle interactions scatter the particles and drive
the CR population towards isotropy in wave frame. 
As a result, it slows down and eventually reverses the growth of $v_s^{\rm eff}$.
The overshoot of $v_s^{\rm eff}$ is an inevitable consequence of this initial relaxation
process, reaching a maximum value of about $15v_A$.
Seen in Figure \ref{fig:fpar_hist}, for particles with momentum around $p=p_0$, level
of anisotropy is the highest around time $2\times10^4\Omega_c^{-1}$, and it is
substantially reduced by the time of $5\times10^4\Omega_c^{-1}$.

After about $5\times10^4\Omega_ct$, a quasi-stationary state is established, where
$v_s^{\rm eff}$ about stays at a constant of $\sim4v_A$, with wave growth and IN
damping balancing each other.
The system still evolves afterwards, with wave energy very slowly increasing over time,
accompanied by a slow reduction of $v_s^{\rm eff}$. This is a result
of the development of the CRSI towards longer wavelengths, together with the
response from more energetic particles that they resonate with. We see from Figure
\ref{fig:fpar_hist} that $\delta f/f_0$ has hardly changed for particles near momentum
$p\sim p_0$, but the level of anisotropy is further reduced for particles with $p\sim10p_0$.

Next, we examine the bulk properties of the CRs and waves in the spatial domain in
Figure \ref{fig:CRWv_prof}. As discussed in Appendix \ref{app:frame}, there can be some
small mismatch at the boundaries, and the CR pressure gradient in the box does not
necessarily match what we impose. More quantitatively, the gradient is reflected in
the value of $L_{\rm CR}$ from Table \ref{tab:params}. For the fiducial run, our imposed
gradient would yield $L_{\rm CR}=2.5\times10^7$, and we see that the actual gradient
in the box center is slightly steeper, with deviation less than $10\%$.
Similarly, the spatial distribution of wave energy
($\int I(k)dk=\langle\delta B^2/B_0^2\rangle$) is not exactly uniform, where it is about
twice in central region than that at the two boundaries. 
For this reason, all our measurements will be conducted
in the central region of the box between the interval $[3\times10^6, 7\times10^6]$ as
marked by vertical dashed lines.

The bottom panel of Figure \ref{fig:CRWv_prof} shows the effective streaming speed
calculated from ${\cal F}_{\rm CR}/(\epsilon_{\rm CR}^{\rm ctr}+P_{\rm CR}^{\rm ctr})$ over the
CR population with $0.1p_0<p<10p_0$.
Note the denominator represents values at domain center (independent of $x$).
The fact that it is flat suggests the system has well arrived at steady state, due to Equation
(\ref{eq:crcont}). 

\begin{figure}
    \centering
    \includegraphics[width=85mm]{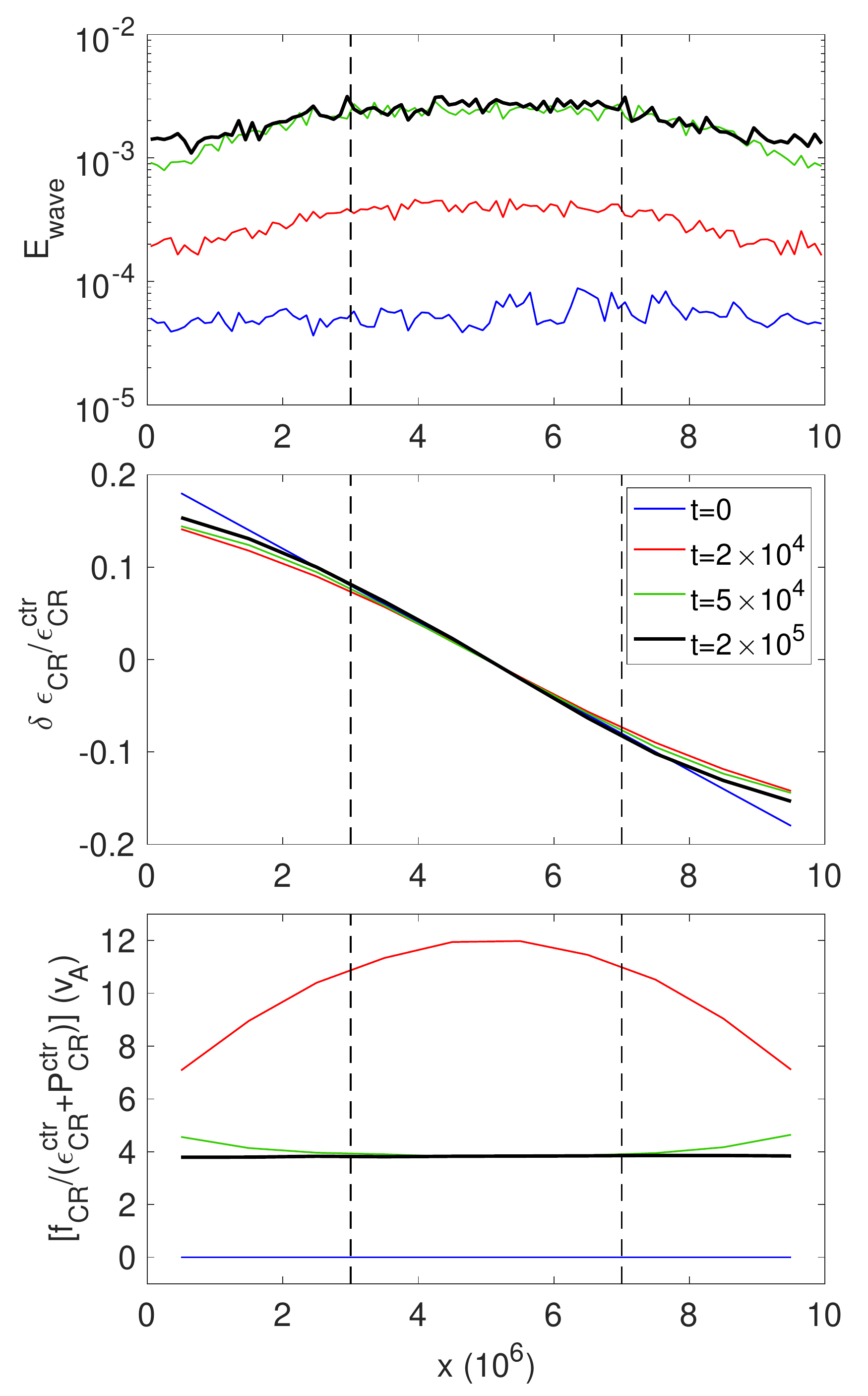}
  \caption{The radial profiles of wave energy density (top), CR energy density
  (deviation from mean, middle), and the CR streaming speed (bottom) from run Fid. 
  In the middle and bottom panels, only CRs with momentum between $0.1p_0$ and
  $10p_0$ are counted. Results are shown at four different snapshots indicated in
  legend.}\label{fig:CRWv_prof}
\end{figure}

\begin{figure*}
    \centering
    \includegraphics[width=180mm]{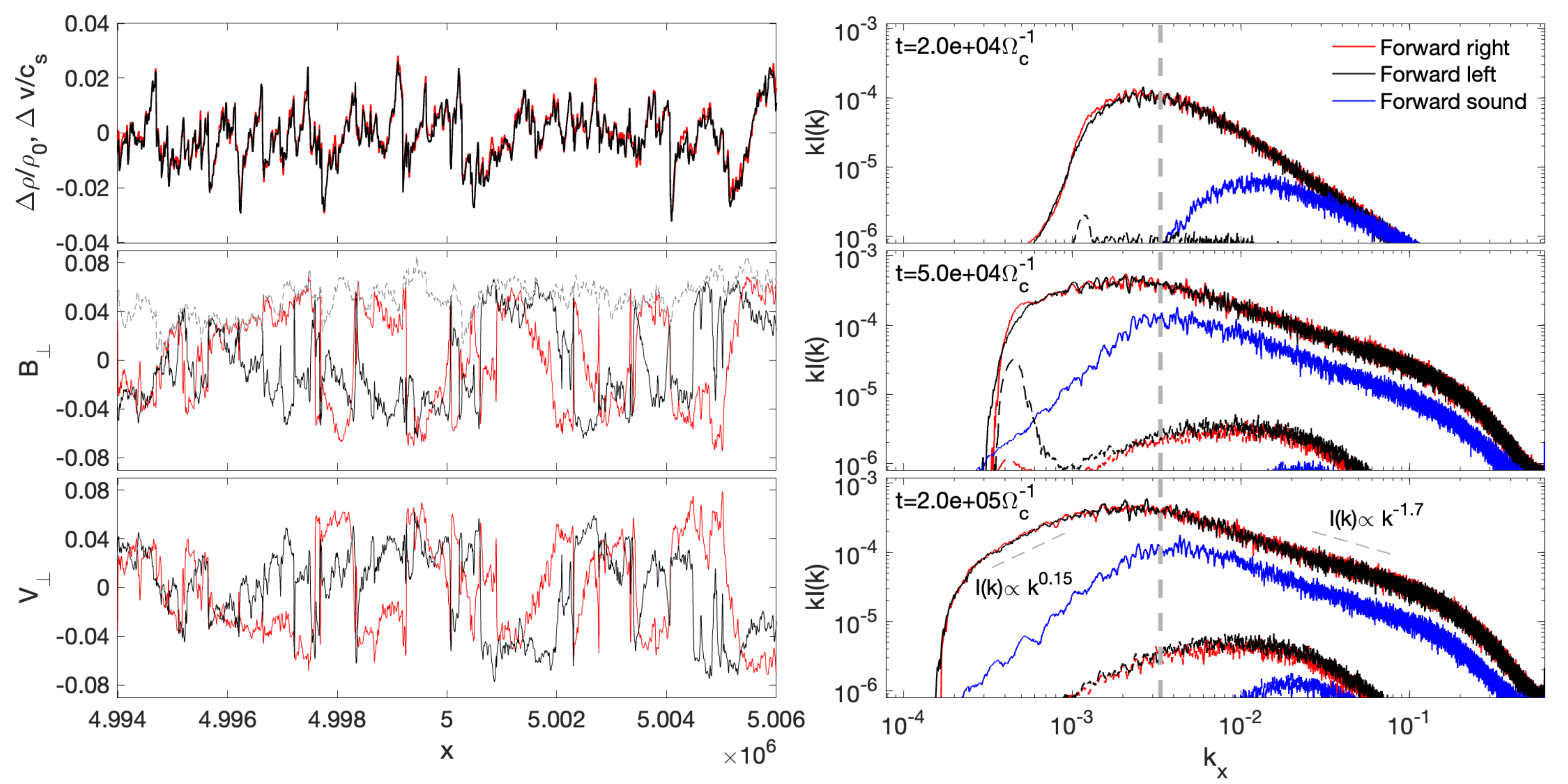}
  \caption{Left: profiles of density (top), longitudinal (top) and perpendicular (bottom)
  velocity and magnetic field (middle) fluctuations in the very central region of the
  simulation box, at the end of the fiducial simulation, showing typical wave patterns.
  Also shown in the middle panel is the magnitude of perpendicular field in grey dashed line.
  Right: spectrum of the waves, shown as dimensionless wave intensity $kI(k)$ at three
  different snapshots (from top to bottom). Waves are decomposed into left/right polarized
  Alfv\'en waves and sound waves. Forward/backward propagating waves are
  shown in solid/dashed lines. The vertical dashed line mark the wave number resonating
  with particles with $p=p_0$ at zero pitch angle.}\label{fig:wave_spec}
\end{figure*}

\subsubsection[]{CR streaming speed}\label{sssec:vdeff}

By examining all simulation runs in the bottom panel of Figure \ref{fig:EwvWeff}, we see
that except for run Nu05, $v_s^{\rm eff}$
converges to almost the same value at saturated state. This is consistent with the expectation
that under linear (ion-neutral) damping, the effective streaming speed should be independent
of the imposed CR pressure gradient. However, there are substantial mismatches from
our theoretical estimates in Section \ref{ssec:theoryall} in two aspects.

First, the converged value of $v_s^{\rm eff}$ of $\sim4v_A$
is larger than the expected value of $1.8v_A$ from (\ref{eq:vsexp}) by a factor of $\sim2$.
The higher value of $v_s^{\rm eff}$ indicates that equating
{\it maximum} growth rate with damping rate, does not lead to a balance, but rather,
modes at almost all wavelengths will be damped.
Therefore, larger $v_s^{\rm eff}$ is needed so that wave amplitudes of a broad
range of wavelengths can be maintained. It needs to be a factor of $\sim2$ when
consider particles in the range of $0.1p_0<p<10p_0$, and the difference would be
even more if extending to higher-energy particles.

Second, by examining the bottom panel of Figure \ref{fig:fpar_hist}, we see that the predicted
momentum-dependent CR streaming speed (\ref{eq:vdprd}) strongly deviates from our
self-consistent measurements: it underestimates $v_s^{\rm eff}$ at $p$ near $p_0$, but
substantially overestimates $v_s^{\rm eff}$ at smaller and larger $p$, even excluding regions
with $p\lesssim0.1p_0$ and $p>10p_0$ not well resolved/accommodated in our simulations.
This is likely the consequence of not capturing the complex wave-particle interaction over
broad wavelengths and particle momenta.

Finally, we observe that the simulation with $\nu_{\rm in}$ reduced by half (run Nu05)
leads to a smaller $v_s^{\rm eff}$ smaller than that in other runs also by half, exactly as
expected.

\subsection[]{Wave spectrum}\label{ssec:spec}

The wave pattern and wave energy spectrum from our fiducial run are shown in Figure \ref{fig:wave_spec}.
The waves are decomposed into left/right polarized Alfv\'en waves following the procedures
in \citetalias{Bai_etal19}, and here we have further incorporated the sound waves. At early time of
$t=2\times10^4\Omega_c^{-1}$, the waves grow the fastest at wave number around
$k_x=k_0=\Omega_c/p_0$, corresponding to wavelength $\lambda_0=2\pi/k_0$.
Both left and right polarized, forward-propagating Alfv\'en waves are excited at the same rate, as
expected. 

Around time $t=5\times10^4\Omega_c^{-1}$, the energy spectrum of the high-$k$ modes
(with $k\gtrsim k_0$) are well established in the form of a power-law with $I(k)\propto k^{-1.7}$,
mimicking a wave cascade that transfers the energy at $k\sim k_0$ to smaller scales.
This process appears strong enough to dominate over the driving from the CRSI itself, and
the overall wave energetics is thus determined by this energy transfer and the ion-neutral damping.
In \citet{Plotnikov_etal21}, we speculated that it
is likely the result of nonlinear steepening of Alfv\'en waves into rotational discontinuities
\citep{CohenKulsrud74}.
Such rotational discontinuities are
readily visible in the middle left panel of Figure \ref{fig:wave_spec} at multiple locations over a short
segment of the simulation box, where one or both of the perpendicular components of the magnetic field
can change abruptly, whereas changes in the magnitude of $B_\perp$ is much more modest at these locations.
We expect it to play a major role helping particles scatter across the $90^\circ$ pitch angle,
as discussed in \citetalias{Bai_etal19}.
The high-$k$ end of the spectrum is subject to numerical dissipation, corresponding to the transition
to steeper slope at around $k\sim0.2$, or a wavelength of $\sim\lambda_0/60$ (resolved by $\sim6-7$
cells). We will see that this level of resolution
is generally sufficient for convergence for the bulk particle population with $p\gtrsim p_0$.

At later time, lower-$k$ modes gradually catch up, as CRSI modes grow more slowly towards longer
wavelengths. Note that by showing the dimensionless $kI(k)$, the lines in the right panels
of Figure \ref{fig:wave_spec} directly indicate the energetics of individual modes. The spectrum
peaks between $k\sim(0.5-1)k_0$, and drops towards longer wavelengths mainly as a result
of fewer CRs energetic enough to resonantly interact with these waves.
The system develops a different power law with $I(k)\propto k^{0.15}$ over a relatively
short range between $k_x\sim0.1-0.5k_0$. 

In steady state, the wave spectrum rapidly drops and cuts off at
$k\lesssim k_{\rm min}=2\times10^{-4}$, or $\sim0.06k_0$. This cutoff
is physical due to ion-neutral damping. As a result, there is not much
isotropization for particles with $p\gtrsim10p_0$. To examine this, we may set 
$v_s^{\rm eff}\sim15v_A$ as mean value for particles with $p\gtrsim10p_0$, as seen from
the bottom panel of Figure \ref{fig:fpar_hist}. By requiring
$\Gamma_{\rm CR}>\nu_{\rm in}/2=0.5\times10^{-4}$, we find $k\gtrsim0.06k_0$, which
agrees well with the cut-off observed in the wave spectrum.

Interestingly, we also observe the growth of forward-propagating sound waves, initially at high-$k$
and later extending to lower-$k$ covering most of the wavelengths where the CRSI operates. The
wave energy spectrum of the sound waves run parallel with the Alfv\'en waves at high-$k$ while being
lower by a factor of $\sim2$, indicative of similar origin (also present though not discussed in our
previous studies, in \citetalias{Bai_etal19}, \citealp{Plotnikov_etal21,Bambic_etal21}). It is also unsurprising,
as the CRSI primarily excites linearly-polarized Alfv\'en waves (left/right modes share similar amplitudes)
that lead to variations in perpendicular magnetic pressure. 
We anticipate that the sound waves have minimum influence to particle scattering on our simulations
(which only introduces higher-order corrections), but multi-dimensional simulations are needed to
further examine its impact. Especially, the resulting oblique magnetosonic modes would enhance
the mirror effect as another important mechanism for scattering particles across the $90^\circ$ pitch
angle barrier \citep{FeliceKulsrud01,HolcombSpitkovsky19}.

\subsection[]{CR scattering}\label{ssec:crscat}

The main goal of this work is to provide self-consistent measurements of the CR scattering rates
from our first-principle simulations, and here we lay out the procedures.

\begin{figure*}
    \centering
    \includegraphics[width=88mm]{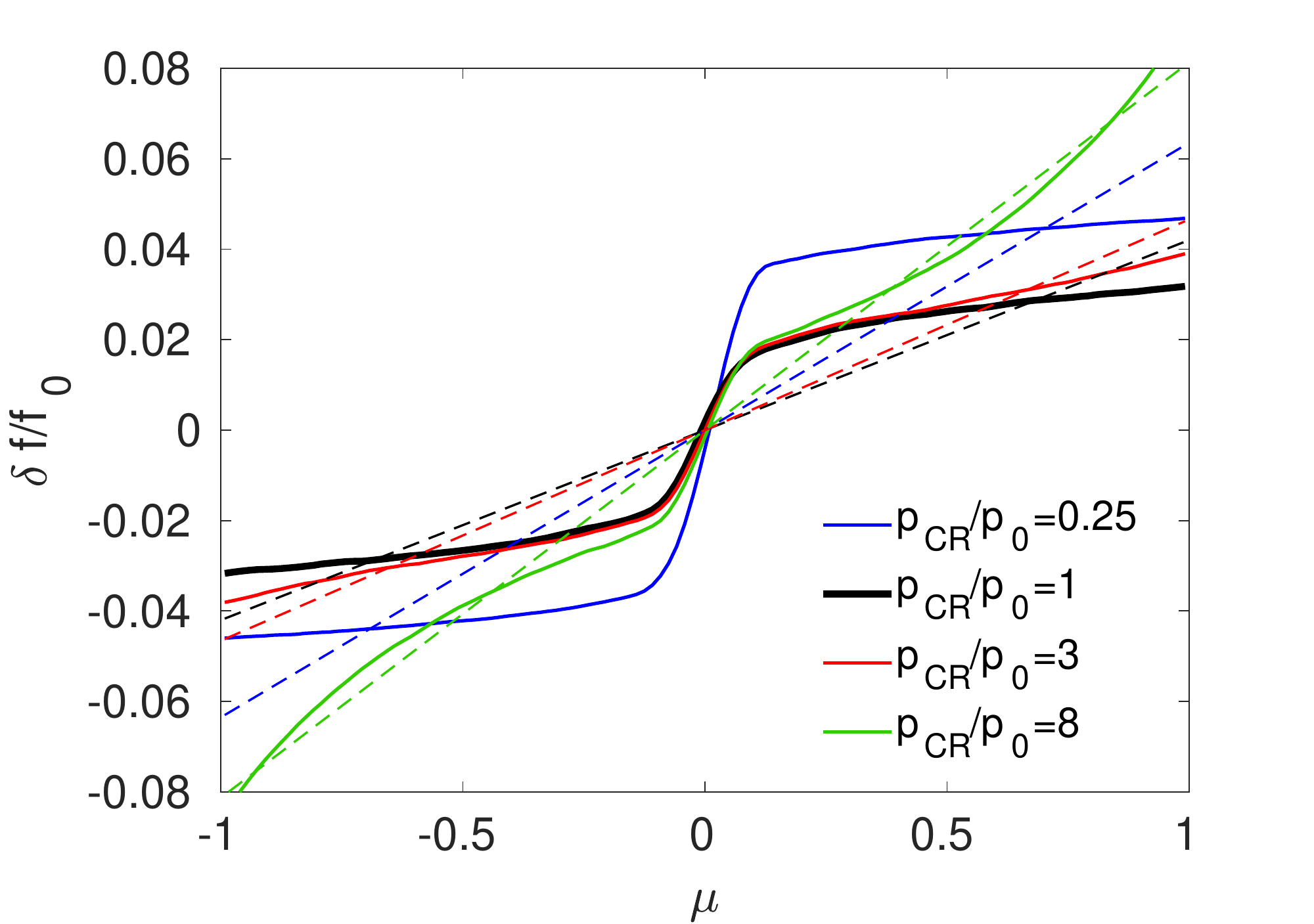}
    \includegraphics[width=88mm]{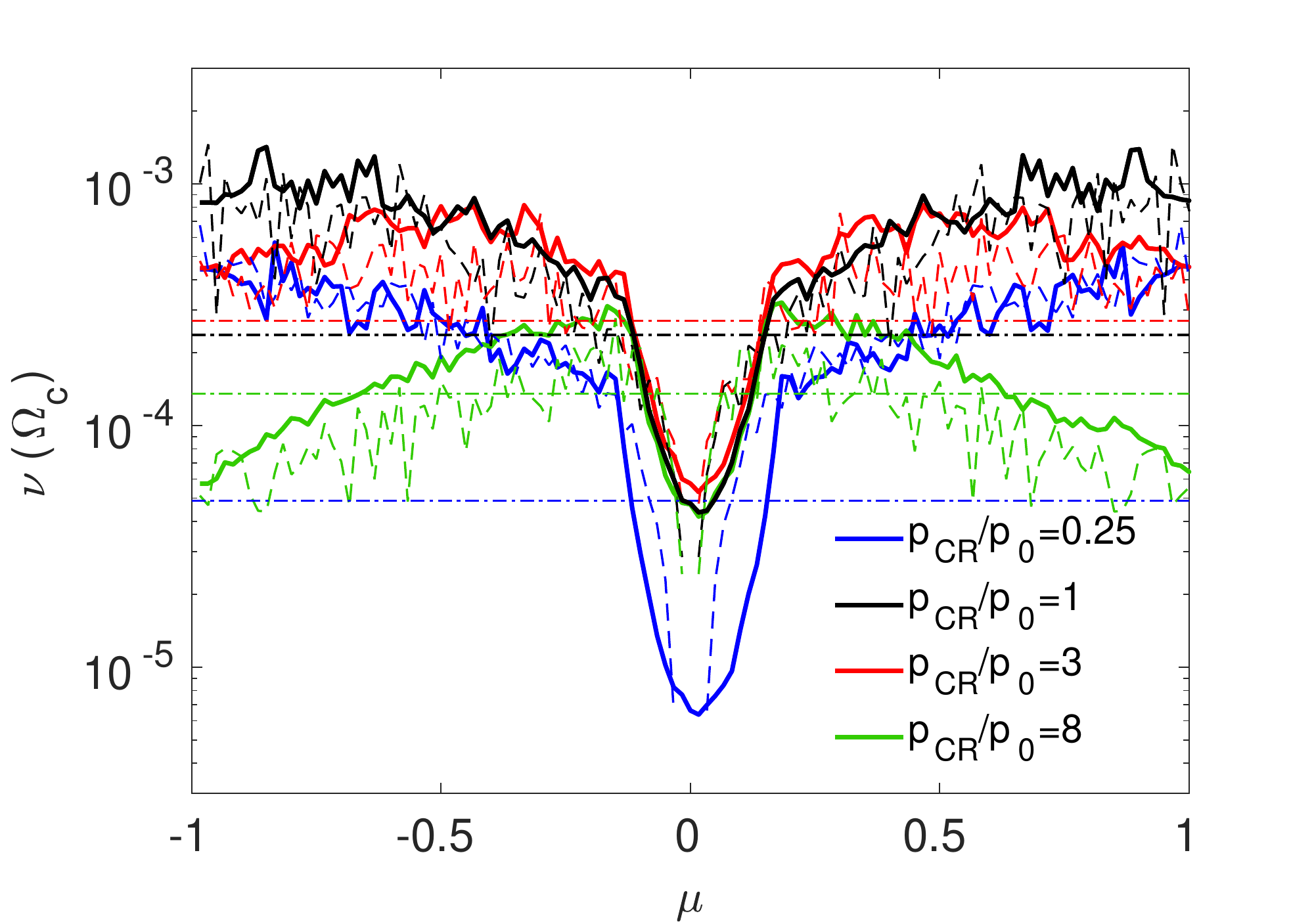}
  \caption{Left: deviation of the particle distribution function $f(p,\mu)$ from $f_0(p)$,
  shown at given $p=0.25p_0, p_0, 3p_0$ and $8p_0$ (with different colors)
  as a function of pitch angle cosine $\mu$. The corresponding dashed lines are line
  fits under the Eddington approximation (\ref{eq:edddef}).
  Right: the solid lines show the resulting total CR scattering rate obtained from Equation (\ref{eq:nu_ss})
  for particles with the same momenta shown on the left. The dashed lines show the expected scattering
  rates based on QLD, given in Equation (\ref{eq:nu_qld}). The horizontal dash-dotted lines indicate the
  effective scattering rate for particles of the given momenta, given by Equation (\ref{eq:nueff}).}\label{fig:nuofmu}
\end{figure*}

\subsubsection[]{Pitch angle distribution}

In Figure \ref{fig:nuofmu}, we show in the left panel the pitch angle distribution at four representative
particle momenta, taken from the fiducial run. This result can be interpreted from the wave-frame
Fokker-Planck equation discussed in Section \ref{ssec:FP}. In particular, the right-hand-side of
Equation (\ref{eq:nu_ss}) is independent of $\mu$, which can be directly evaluated at the center of
our simulation box. Therefore, the slope of $d\delta f/d\mu$ is inversely proportional to the pitch
angle scattering rate $\nu(p,\mu)$, and it also reflects the level of isotropization at given $\mu$.

For these particles with representative momenta, full isotropization in the wave frame is never
achieved. For particles with $p\lesssim3p_0$, the DF is close to isotropic (i.e., curve close to being flat)
when $\mu$ is away from $0$, indicating efficient scattering, but develops a steep slope around
$\mu=0$. This is a manifestation of the well-known $\mu=0$ barrier, or $90^\circ$ pitch angle barrier,
where QLT fails. The fact that the slope is not infinite indicates that non-linear effects contribute to
scattering.
In fact, we have analyzed individual particle trajectories to examine the source of nonlinearity, and
the results are very similar to \citetalias{Bai_etal19} (and hence not shown here),
where reflections are generally found to be associated with particles
entering the rotational discontinuities, a manifestation of non-linear wave-particle interaction.
For particles with $p=8p_0$, the fall-off of low-$k$ modes makes their isotropization less efficient,
leading to an increase of the $df/d\mu$ slope towards $|\mu|=1$, whereas the slope near $\mu=0$
remains similar to that for lower-energy particles.

We further examine the validity of the Eddington approximation (\ref{eq:edddef}), which essentially
corresponds to fitting the $f(p,\mu)$ curve as a function of $\mu$ by a straight line. The results are
shown as dashed lines
in the left panel of Figure \ref{fig:nuofmu} for each momentum. Generally, we see substantial
deviations near $\mu=0$ because of the steepening of the $df/d\mu$ slope. Such deviations
undermine the accuracy of this approximation. In particular, from Equations (\ref{eq:df1}) and
(\ref{eq:fluid1}), the effective scattering rate is weighted by the $df/d\mu$ slope, and hence the
less-isotropized $\mu\approx0$ region carries a factor of a few more weight,
a fact that was also noted in \citet{Bambic_etal21}. As this region is
characterized by smaller scattering rates, we thus anticipate the overall effective scattering rates
to be smaller than those obtained by assuming Eddington approximation.

\subsubsection[]{Primitive scattering rates}\label{ssec:primsca}

In the right panel of Figure \ref{fig:nuofmu}, we show the inferred full scattering rates
$\nu(p,\mu)$ from Equation (\ref{eq:nu_ss}). For comparison, we also show the
expected scattering rate from QLD (\ref{eq:nu_qld}). We note that in computing these
quantities, the derivative $df/d\mu$, as well as the raw wave energy spectrum, can be very
noisy. We employ a Savitzky-Golay filter of with $n_L=n_R=6$, $M=2$, applied to both
$f(p,\mu)$ and $I(k)$, to help mitigate the noise (which is also applied to obtain the right
panel of Figure \ref{fig:wave_spec}).

Generally, for particles with $p\lesssim3p_0$, we find good agreement between the
derived $\nu_\mu(p,\mu)$ and $\nu_{\rm QL}(p,\mu)$ when $\mu$ is away from $0$. This
is consistent with that obtained in \citetalias{Bai_etal19}, where detailed comparison
between QLD and particle pitch angle evolution was carried out. At $\mu\approx0$,
on the other hand, we see that scattering rate from the QLD falls towards 0, whereas
$\nu(p,\mu)$ remains finite, thanks to contributions from non-linear effects that assist
the particles overcoming the $\mu=0$ barrier.

\begin{figure*}
    \centering
    \includegraphics[width=85mm]{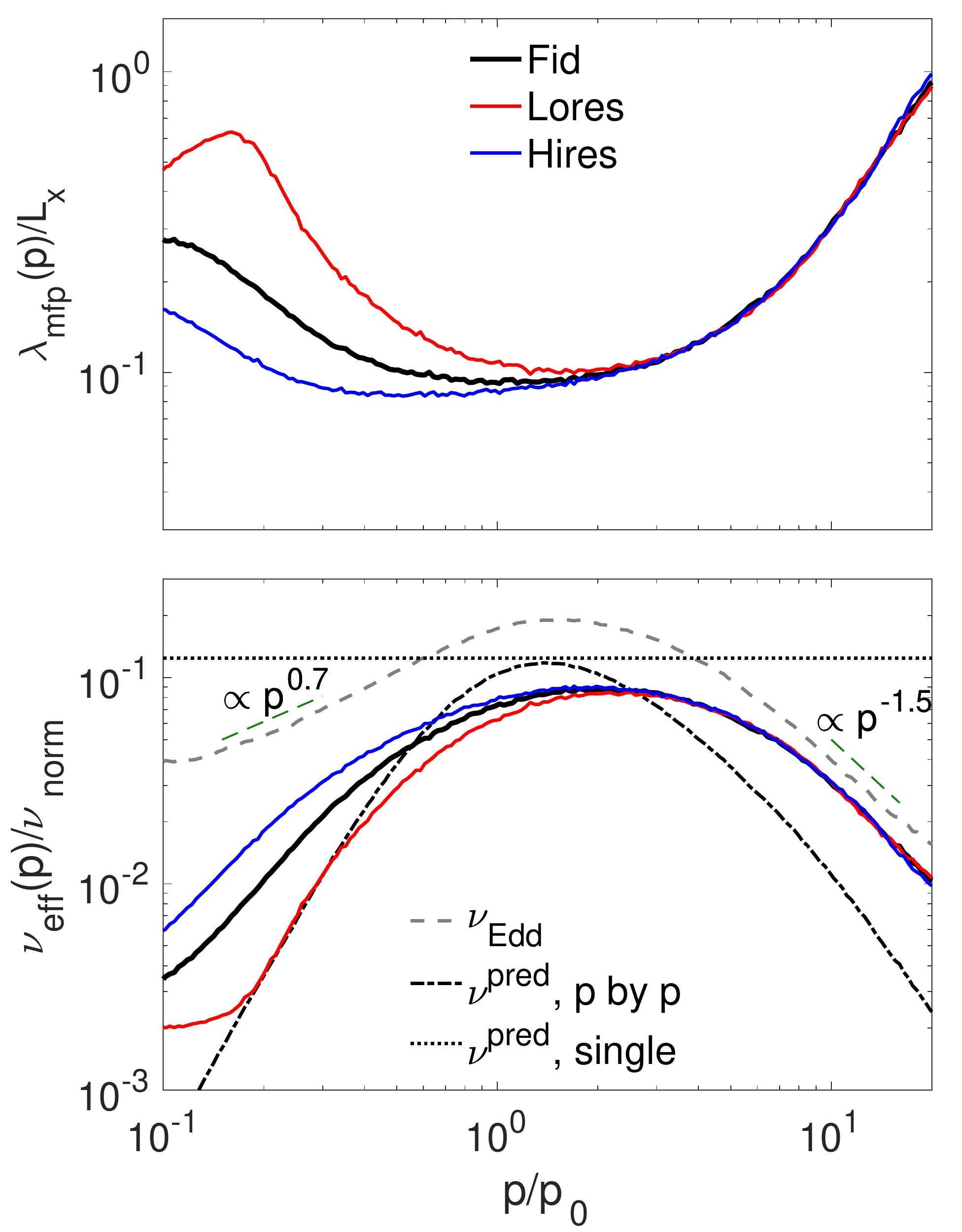}
    \includegraphics[width=85mm]{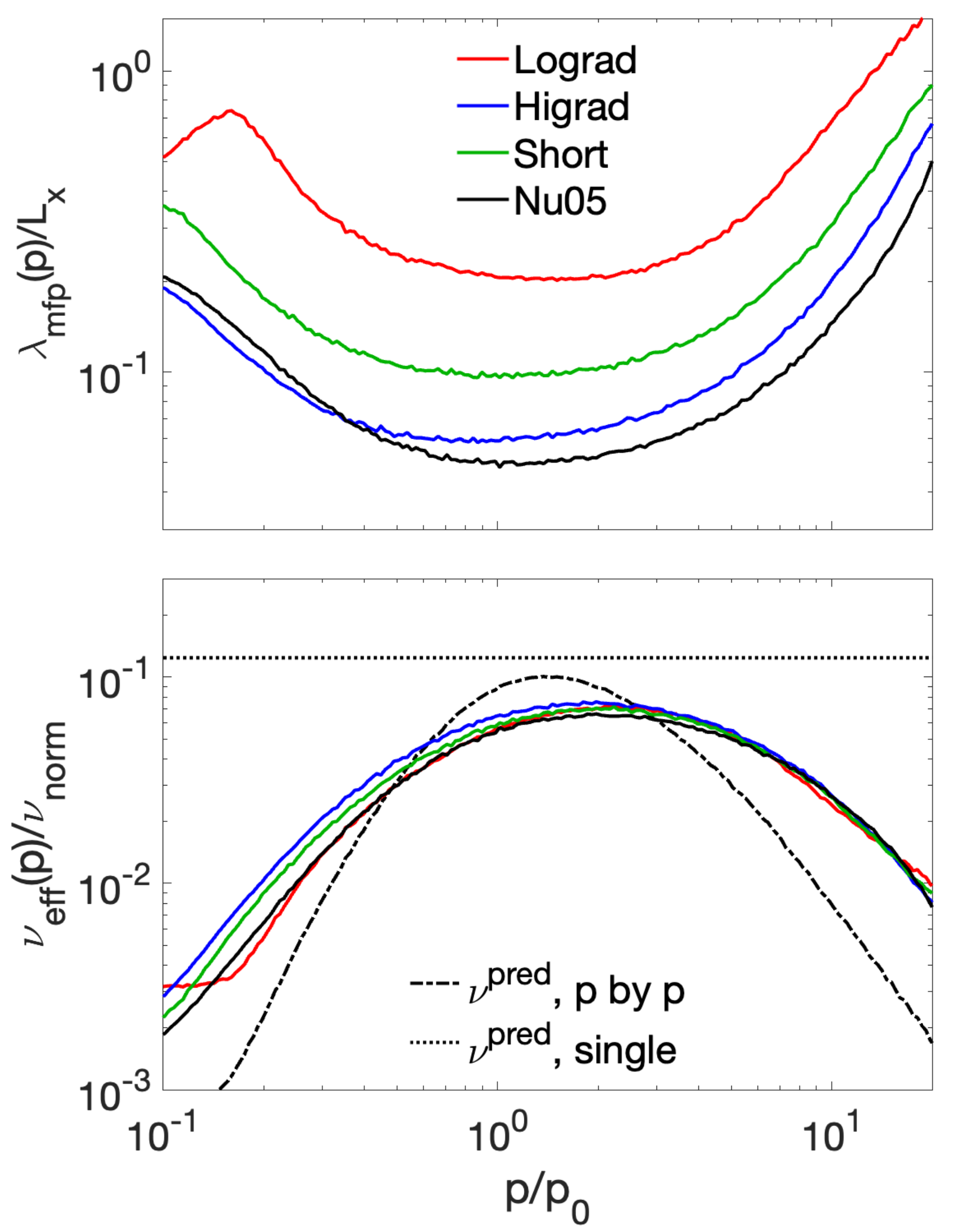}
  \caption{The effective scattering rate $\nu_{\rm sca}(p)$ normalized by $\nu_{\rm norm}$ (bottom),
  and the particle mean free path $\lambda_{\rm mfp}(p)$ normalized by simulation box size $L_x$ (top),
  shown as a function of momentum.
  Results are compared for runs with different resolutions (left) and with different CR pressure gradient
  and ion-neutral damping rate (right). In the bottom panels, we further show the theoretical scattering
  rates based on momentum-by-momentum treatment (\ref{eq:nuscapred}) in dash-dotted lines, as well
  as from single-fluid treatment (\ref{eq:nuscaeff}) in horizontal dotted lines.
  Also shown in dashed line in the bottom left panel are the scattering rates expected from the Eddington
  approximation (\ref{eq:eddasump}) based on our measured wave spectrum.}\label{fig:nuofp}
\end{figure*}

For particles with $p=8p_0$, we find our derived $\nu_\mu(p,\mu)$ is slightly higher
than predicted from QLD. We have further examined particles with higher momentum,
and find that deviation gradually increases.
The underlying reason is that these particles are not sufficiently strongly
scattered with a mean free path $\lambda_{\rm mfp}$ not small compared to simulation
box size $L_x$ (see Figure \ref{fig:nuofp} and Section \ref{ssec:nuofp}).
Our simulation time is not yet long-enough to fully relax the spatial gradients,
leading to an overestimate of $\pa P_{\rm CR}/\pa x$.\footnote{Consider the case without scattering,
one would eventually arrive at $\pa P_{\rm CR}/\pa x=0$. However, as we imposed a non-zero
$\pa P_{\rm CR}/\pa x$ at $t=0$, reducing the gradient to zero would require all initial particles to
traverse the box. For particles with pitch angles near $\mu=0$, this would
take many times $L_x/{\mathbb C}$, which is well beyond our simulation time.}
Nevertheless, the deviation is within a factor of two for $p\lesssim10p_0$ and the
effect on the overall momentum-integrated fluid scattering rate (to be discussed in
Section \ref{ssec:nutot}) is very minor.

Integrating over $\mu$, or from Equation (\ref{eq:nueff}), we further show in horizontal
dash-dotted lines the value of $\nu_{\rm sca}(p)$ at individual momentum $p$.
This value is self-consistently measured from $F_0$ and $F_1$,
reflecting the aforementioned weighted average in Equations (\ref{eq:df1}) and
(\ref{eq:fluid1}). We can see by eye that the level indicated by $\nu_{\rm sca}$ is
lower compared to a simple average over the curve, and this is the most obvious
for low-energy particles ($p\lesssim p_0$), highlighting the importance
of kinetic simulations capturing the physics of CR scattering near $\mu=0$.

While not shown here, we have also tracked the trajectories of a small fraction of
particles to directly evaluate $D_{\mu\mu}$ defined in (\ref{eq:Dmumu}). The results
are also in general agreement with our measured results.
We note that similar agreement has been found in \citet{Bambic_etal21}.

\section[]{Simulation results: Fluid Scattering Rates from All Runs}\label{sec:param}

In this section, we compile data from all simulations and evaluate the
momentum-by-momentum fluid scattering rate $\nu_{\rm sca}(p)$ from Equation
(\ref{eq:nueff}), from which we can also obtain particle mean free path $\lambda(p)$.
The results are shown in Figure \ref{fig:nuofp}, with $\nu_{\rm sca}(p)$ normalized by
$\nu_{\rm norm}$ (\ref{eq:nu_norm}), and $\lambda_{\rm mfp}$ normalized by $L_x$.
For comparison, we also show the predicted scattering rates in dotted and dash-dotted lines
for single-fluid and momentum-by-momentum treatments. 
The total fluid scattering rate $\nu_{\rm sca}$ is then obtained by a weighted average
of $\nu_{\rm sca}(p)$ over momentum with $0.1p_0<p<10p_0$, given by Equation (\ref{eq:nufluid}).
Major diagnostic quantities associated with all simulation runs are also listed in Table
\ref{tab:params}, including the wave amplitudes $E_{\rm wave}$,
$L_{\rm CR}$ at the simulation box center, and the fluid scattering rate coefficient $\alpha_{\rm sca}$.

\subsection[]{Scattering rates and mean free path by momentum}\label{ssec:nuofp}

As a function of momentum, we see from the bottom left panel of Figure \ref{fig:nuofp} that
$\nu_{\rm sca}(p)$ is maximized around $p\sim p_0-4p_0$, and falls off on both ends.
Simulations with different resolutions agree for $p\gtrsim 2p_0$. In the low-$p$ end,
higher resolution yields higher scattering rates, and convergence is reached between run
Fid and Hires down to about $0.5p_0$. This is a consequence of higher resolution being
capable of better resolving smaller scales from the $I(k)\sim k^{-q}$ spectrum.

In the low-$p$ range, we notice from the highest resolution run that at $p\lesssim p_0$, a
power-law is developed with $\nu_{\rm sca}\propto p^{0.7}$. The slope of $-0.7\approx-(q-1)$
for non-relativistic particles is in line with the scaling from the Eddington approximation (\ref{eq:nu2}).
In fact, the expected scattering rate from the Eddington approximation (\ref{eq:eddasump}) based
on our measured wave spectrum, shown in dashed line for run Fid, just has a normalization factor
$\sim2.5$ times higher (as it does not account for inefficient scattering across the $\mu=0$ barrier).
Therefore, even without reaching full convergence down to small $p$, we can well
predict the converged behavior, and results from highest resolution is in fact close to this
expectation down to $\sim0.3p_0$. On the other hand, the predicted
$\nu_{\rm sca}^{\rm pred}(p)$ for $p\lesssim p_0$ has a very different slope $\propto p^3$,
leading to substantial deviations in the low-$p$ end. This is the direct consequence of the
$I(k)\sim k^{-1.7}$ spectrum discussed in Section \ref{ssec:spec} that overcomes wave damping
and hence is much harder than standard expectations.

In the high-$p$ range,  we find a power law of $\nu_{\rm sca}\propto p^{-1.5}$ which is, surprisingly,
in line with the predicted slope of $p^{1-2\kappa}$ even though the low-$k$ wave spectrum
relevant to the resonant scattering does not show an extended power law. On the other hand, due
to the cutoff in low-$k$ wave spectrum, we expect the scaling of $\nu_{\rm sca}(p)$ with $p$ to drop
off beyond $p\sim10p_0$, though it is not yet seen due to limited domain size and simulation time. 
Our measured scattering rates are higher then predicted by a modest factor of $2-3$.

The top left panel of Figure \ref{fig:nuofp} shows the particle mean free paths $\lambda_{\rm mfp}$
as a function of $p$. Given our simulation box size of $L_x=10^7d_i$, we see that
$L_x\gtrsim5\lambda_{\rm mfp}$ for particles with $p\lesssim6p_0$, ensuring enough scattering
within the simulation box to validate our streaming box setting. We note that despite of having
$\nu_{\rm sca}$ decrease with momentum for low-energy particles, their $\lambda_{\rm mfp}$
remains close to flat. This is because these particles are sub-relativistic, and the reduction of particle
speed largely cancels the reduction of $\nu_{\rm sca}$ (see Equation (\ref{eq:nueff})). For more
energetic, relativistic particles, on the other hand, their $\lambda_{\rm mfp}$ increases rapidly with
$p$, associated with the reduction of $\nu_{\rm sca}(p)$ in the high-$p$ end.

We further examine the role of CR pressure gradient and ion-neutral damping rate in the right panels
of Figure \ref{fig:nuofp}. Most features are similar to those in the fiducial run.
Remarkably, the measured scattering rates for all these runs, when normalized by $\nu_{\rm norm}$,
largely overlap with each other, and they deviate from the predicted scattering rate in the same way
as discussed earlier for the fiducial run. This is strong indication that the physics of the CRSI is
independent of these parameters that we have varied, as anticipated from QLT.
We have also examined the wave spectrum and found that their
shapes are all similar to that of the fiducial run in both the high-$k$ and low-$k$ ranges. Besides, the
run with higher/lower CR pressure gradient or weaker/stronger damping rate shows a
lower/higher $k_{\rm min}$ in proportion with the gradient, in line with the discussion in Section
\ref{ssec:spec}. For most runs, the particle mean free path for the bulk CR population is well within
our simulation box size. The case of run Lograd with relatively small imposed CR pressure
gradient is more marginal, but still guarantees $L_x\gtrsim3\lambda_{\rm mfp}$ for
$0.4p_0\lesssim p\lesssim5p_0$. 

Overall, the predicted scattering rate based on single-fluid treatment (\ref{eq:nuscaeff}) is always
higher than the measured growth rate by a factor of $\sim2$ at the peak with more significant
overestimates at other momenta.
The momentum-by-momentum treatment (\ref{eq:nuscapred}) yields closer estimate near the peak,
though it significantly underestimates the scattering rates at the low-$p$ end, and it also
underestimates the scattering rates at the high-$p$ end by a factor of a few.

\begin{figure}
    \centering
    \includegraphics[width=85mm]{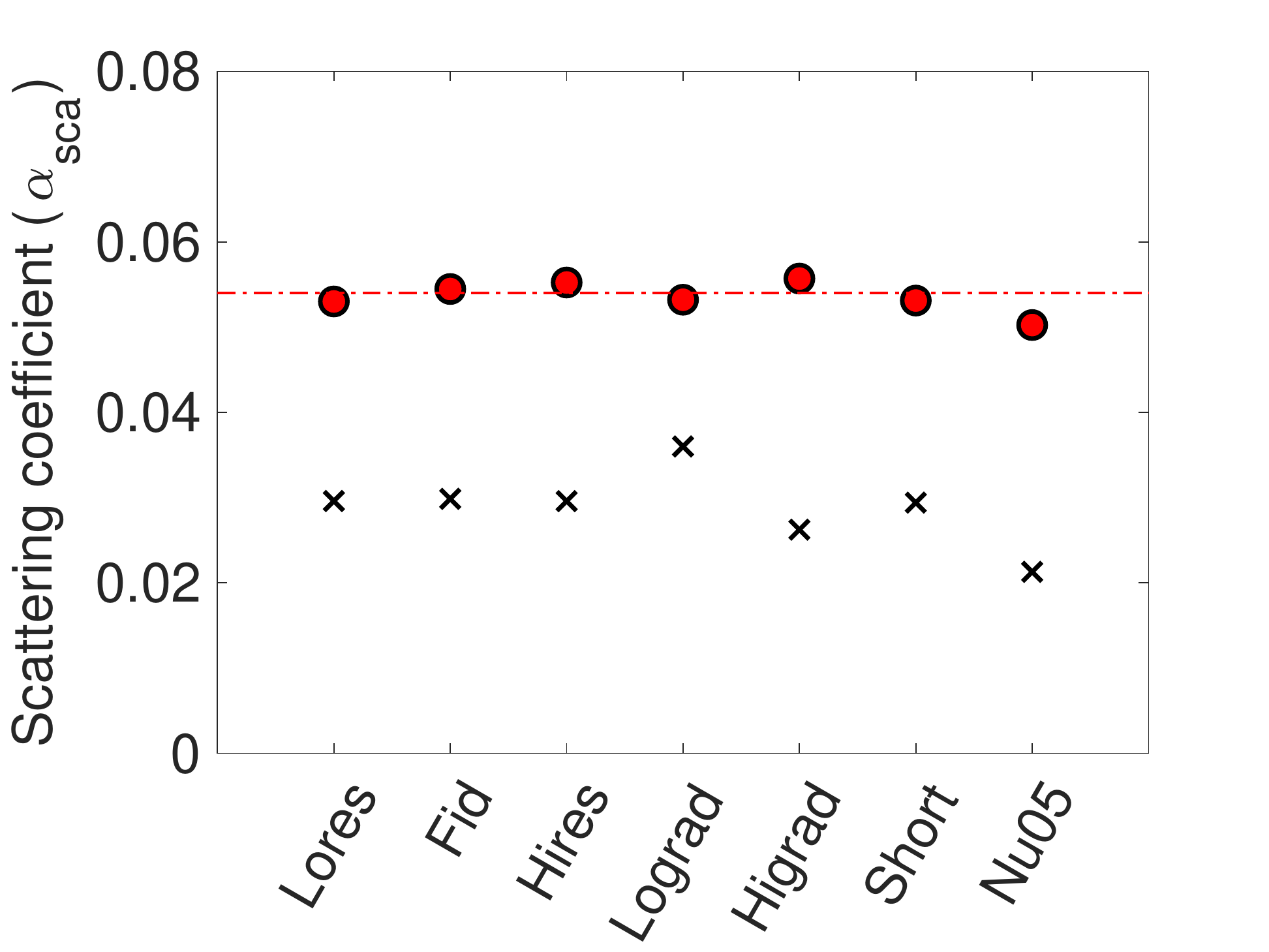}
  \caption{
  Measured dimensionless fluid scattering rates from
  all our simulation runs, normalized by $\nu_{\rm norm}$ given by (\ref{eq:nu_norm}).
  Red filled circles correspond to results obtained by integrating over
  finite momentum range $0.1p_0\leq p \leq 10p_0$, and black crosses correspond to those
  integrated over full range.}\label{fig:nu_summary}
\end{figure}

\subsection[]{Total fluid scattering rates}\label{ssec:nutot}

Finally, we discuss the fluid scattering rate $\nu_{\rm sca}$, computed from a weighted
average of $\nu_{\rm sca}(p)$ over $p$ given by Equation (\ref{eq:nufluid}). The results are
normalized by $\nu_{\rm norm}$, with dimensionless scattering coefficient $\alpha_{\rm sca}$
measured from all runs shown in Figure \ref{fig:nu_summary}.
As particles with $p\lesssim0.1p_0$ and $p\gtrsim10p_0$ are not very effectively scattered
due to the lack of resolution (low-$p$) and finite box size and simulation time (high-$p$), we
report the main results obtained by integrating over $0.1p_0\leq p\leq p_0$ (in red filled circles).
For comparison, we also show results obtained by integrating over the full range of $p$
(in black crosses). In the measurements, we use the value of $L_{\rm CR}$ measured directly
from the simulations, given in Table \ref{tab:params}. We note that the values of $L_{\rm CR}$
typically deviate from $(n_{\rm CR}^{\rm ctr}/\Delta n_{\rm CR})L_x$ by about $\lesssim10\%$ (except for run
Lograd where deviation is $18\%$). This correction is
important given the level of precision in our measurements.

Remarkably, the numerical coefficients from all runs, regardless of the imposed pressure
gradients and damping rates, agree very well with each other. Given the same resolution,
runs with CR pressure gradient varied by more than a factor of $4$ yield values that differ
by no more than $5\%$. This result firmly establishes the scaling anticipated from QLT, and
our work provides the accurately calibrated scattering coefficient. For particles with
$0.1p_0\leq p\leq 10p_0$, we find
\bgeq\label{eq:nuscafluid}
\nu_{\rm sca}\approx0.054\nu_{\rm norm}\ .
\edeq

If we consider the full range of particle momenta, as shown in black crosses in Figure
\ref{fig:nu_summary}, we see that the coefficient $\alpha_{\rm sca}$ is further lowered by a
factor of $\sim2$ to $\alpha_{\rm sca}\sim0.03$. The lowered values are almost entirely due
to particles with momenta $p>10p_0$ that we do not well accommodate. In Section
\ref{ssec:primsca}, we have discussed that for such particles, due to limited simulation time,
our estimates of $\nu_{\rm sca}(p)$ is higher than the true value by a factor of
$\gtrsim2$. Therefore, if we extend $p_{\rm max}$ to $100p_0$, we expect the true
value of $\alpha_{\rm sca}$ to be even smaller, which we may quote
$\alpha_{\rm sca}\approx0.01-0.02$ as the final scattering coefficient inferred from our simulations.

The coefficient $\alpha_{\rm sca}\approx0.054$ or $0.01-0.02$, is smaller by a factor
of $\sim2.5-10$ compared to our predicted value of $\sim0.124$ when treating
CRs as a single fluid. This reduction is mainly due to the reduction of $\nu_{\rm sca}(p)$ (thus
enhancement of $v_s^{\rm eff}$) compared to the Eddington approximation, and accurate
weighting from the momentum integral.
On the other hand, our measured $\alpha_{\rm sca}$ coefficient is
comparable to the value of $0.046$ or $0.0064$, obtained from a momentum-by-momentum
treatment mentioned at the end of Section \ref{sssec:nu_mom}.
This agreement reflects some lucky cancellations: the predicted scattering rate is
higher at the peak momentum, but lower at the low-$p$ and high-$p$ ends. For practical
purposes, it is encouraging and suggests that
Equations (\ref{eq:nuscapred}) and (\ref{eq:vscprd}) are a viable prescription of 
CR scattering rates for $p\gtrsim p_0$.

\section[]{Discussion}\label{sec:discussion}

\subsection[]{Comparison to previous kinetic simulations}

As a first kinetic study of the CRSI with driving thanks to the streaming box framework, our
results can be compared with previous kinetic simulations with periodic boundary conditions
(e.g., \citealp{Bai_etal19,HolcombSpitkovsky19,Plotnikov_etal21}).
These simulations rely on a pre-existing CR streaming speed to excite the instability, and
die out as the free-energy from initial CR streaming is exhausted. Usually, a large initial
streaming speed or large CR number density drives waves to larger amplitudes, and
facilitates particles crossing the $\mu=0$ barrier, leading to full isotropization in the wave
frame. On the other hand, with (more-realistic) low wave amplitudes, particles are isotropized
separately for $\mu>0$ and $\mu<0$ regions, with a discontinuous jump at $\mu=0$ owing to
the $90^\circ$ barrier, and the system reaches pre-matured saturation without full isotropization.
We see from our new simulations that in the presence of background CR pressure gradient
mimicking more realistic situations,
the effective CR streaming speed is determined and maintained self-consistently, and by
balancing growth with damping, steady wave amplitudes are achieved to allow CRs to develop
a steady DF. We see that CRs are neither fully isotropized in the wave frame (which
would reduce streaming speed in wave frame to zero), nor get stuck at $\mu=0$ (the DF is
steepened but without discontinuous jump at $\mu=0$).

Even in a periodic box, a CR pressure gradient can be achieved by introducing an artificial interface
separating regions with and without damping, thus breaking the translational symmetry. This
approach allows \citet{Bambic_etal21} to make an attempt to compare kinetic simulation results to
fluid theory over a transient phase, finding good agreement. Because the simulations are also
powered by initial CR streaming, the comparison with CR fluid theory includes a time-dependent
term which accounts for the reduction of CR streaming over time.
It could serve as a first step towards kinetic studies of multi-phase media, and we anticipate that
our streaming box framework can also be  extended to such studies to allow for a steady-state
characterization of CR propagation in multi-phase media.

\subsection[]{Implications for CR magnetohydrodynamics}

A major goal of our study is to provide self-consistent scattering coefficients as sub-grid models for
CR-MHD. While the CRs in CR-MHD would be dynamically-evolving, the steady state
achieved in our kinetic simulations are valid and well justified to apply because of the dramatic
separation between kinetic and fluid scales. The timescale for fluid environment to vary is expected
to be on the order of $L_{\rm CR}/v_A$, and the timescale for CR response is $\sim\nu_{\rm sca}^{-1}$.
Using (\ref{eq:nuscafluid}), the ratio of the two is given by
$L_{\rm CR}\nu_{\rm sca}/(v_A)\approx0.03(n_{\rm CR}/n_i)(c/v_A)^2(\Omega_c/\nu_{\rm in})$,
which is about $3\times10^{3}$ in our simulations. In more realistic environment, while there are
multiple factors contributing, we anticipate $(n_{\rm CR}/n_i)(c/v_A)^2\sim$ order unity
in typical Galactic ISM conditions, and $\nu_{\rm in}$ being orders of magnitude lower than
$\Omega_c$, thus we expect the ratio to be $\gg1$ in general, justifying the required
scale-separation.

So far, most studies of CR hydrodynamics and CR transport adopt simple prescriptions of CR
scattering rates or CR diffusivity, mostly as a constant. Very recently, more realistic
prescriptions of CR transport are starting to be employed, based on QLT estimates
\citep{Hopkins_etal21b,Armillotta_etal21}. Our momentum-by-momentum formula (\ref{eq:vseff1})
is largely identical to
that adopted by \citet{Armillotta_etal21}, except for the minor difference where we explicitly
treat the $\epsilon_{\rm CR}/(\epsilon_{\rm CR}+P_{\rm CR})$ factor to be
$3{\mathbb C}^2/(3{\mathbb C}^2+v(p)^2)$ as opposed to their $3/4$ assuming $v=c$.
Therefore, the pros and cons discussed in Section \ref{sec:param} well applies to their work.
Although \citet{Hopkins_etal21b} did not consider ion-neutral damping, their approach is very
similar to our single-fluid treatment of CRs, and by the same token, we anticipate their adopted
CR scattering rates can be a factor of $\gtrsim10$ higher than realistic values if they were to
include ion-neutral damping. It is yet to examine how much the results change when
incorporating non-linear damping mechanisms such as non-linear Landau damping. In particular,
as noted earlier, QLT is barely utilized in our derivations of the predicted scattering rates for the
linear ion-neutral damping, and one may expect further deviations given the additional
uncertainties in QLT when considering non-linear damping mechanisms.

We note that \citet{Hopkins_etal21b} calls for reduced CR scattering rates compared to QLT
estimates by a factor of $\sim100$ to match the integrated scattering rates for the Milky Way.
Despite the dominant damping mechanisms in their simulations are not ion-neutral damping,
our results of scattering rate reduction may account for a major fraction of the desired reduction.
On the other hand, by post-processing very high resolution local simulations that well resolve
the ISM, \citet{Armillotta_etal21} concluded there is no single CR diffusivity. Overall, the
consequences of reduced scattering rate compared to those derived from conventional QLT
remain to be seen, but we anticipate the resolution of a major uncertainty in the prescription
will facilitate future studies to incorporate more realistic CR physics.

\subsection[]{Implications for CR transport}

The momentum-dependent CR scattering rates obtained from our simulations provide
important calibrations for understanding CR transport in the self-confinement regime.
Spectral breaks in Galactic CRs at rigidity $R\sim200$GeV have been interpreted as
evidence for CR transport transitioning from being dominated by self-generated waves to
extrinsic turbulence \citep{Blasi_etal12,Evoli_etal18}\footnote{We note that their
adopted wave growth rate, which can be derived by combining our Equations
(\ref{eq:gamcr_pred}) and (\ref{eq:vseff1}) and (crudely) substituting $\nu_{\rm sca}(p)$ by
$\pi\Omega k_{\rm res}I(k_{\rm res})$, is a factor $\sim4$ larger than ours.}. 
Incorporating a well-calibrated CR scattering rate is thus crucial to accurately model CR
transport for CR energies of $\lesssim200$GeV, which also the energy range that plays
a dominant role in CR feedback. 
Recently, it has also become possible to evolve a spectrum of CRs together with MHD
\citep{Girichidis_etal20,Hopkins_etal22a}. Applications of this method to study CR-driven
galactic wind have revealed rich physics and phenomenology, although the studies so far still
assumed constant or phenomenological CR diffusion coefficients \citep{Girichidis_etal21,Hopkins_etal22a}.
We anticipate that our calibrated prescriptions can lead to major improvements for
better fidelity in future simulations.

In the meantime, we note that the standard CR self-confinement models based on QLT has
very recently been called into question, as they appear to fail to reproduce the observed CR spectral
slopes and the spectra of secondary-to-primary CRs (e.g. the B/C ratio), and that self-confinement
may be inherently unstable \citep{KempskiQuataert22,Hopkins_etal22b}. Our simulation results,
being consistent with the QLT scalings, do not directly resolve the issue. However, these works
have not necessarily exhausted all the possibly relevant physics, especially microphysics, and the
nature of the multi-phase interstellar and circumgalactic medium always lead to further
complications. Even though the streaming framework is subject to imposed CR densities at the two
boundaries, it still offers certain freedom to conduct controlled experiments around the box
center region (a few mean free paths away from the boundaries), and thus we anticipate it will
become a powerful tool to further investigate the kinetic aspects of this dilemma.

Besides, low-energy CRs are important source of ionization, especially
in high-density, weakly-ionized regions such as molecular clouds and protoplanetary disks,
which determine how strong gas and magnetic fields couple, with significant dynamical
consequences. There has been considerable recent research interest in studying the generation
and propagation of CRs in these environments (see e.g., \citealp{Padovani_etal20} for a review).
Penetration of CRs in molecular clouds \citep{EverettZweibel11,Ivlev_etal18,SilsbeeIvlev19} plays
a significant role regulating star formation (e.g., \citealp{Zhao_etal20,Semenov_etal20}). The
ionization rate also crucially influences the gas dynamics of protoplanetary disks, where it is
uncertain whether CRs are excluded by stellar and/or disk winds \citep{Cleeves_etal13}. It has also
been proposed that young stellar objects themselves are sources of CRs \citep{Padovani_etal16},
which may significantly impact disk ionization \citep{Offner_etal19,Rodgers-Lee_etal20}.
Due to the largely unknown diffusion coefficients, empirical diffusion coefficients are exclusively
used in model calculations. It is yet to examine whether and where self-generated waves can
dominate CR scattering in such environments, thus helps elucidate and reduce dramatic uncertainties.

\section[]{Summary and Conclusions}\label{sec:conclusion}

In this work, we have described a novel, multi-scale numerical framework, the streaming box, to
simulate the driving of the CRSI at kinetic level and enable self-consistent measurement of the
CR scattering rates. The streaming box can be considered as a local patch along magnetic fields
in a global system, which is macroscopically short compared with system size, but microscopically
long so as to contain multiple CR mean free paths. By imposing and maintaining a CR
pressure gradient at box boundaries, the system develops a CR flux through the box that constantly
drives the CRSI towards steady state, where wave growth is balanced by damping (here
ion-neutral damping), realizing the
classic scenario expected from QLT. This enables us to conduct detailed study to test CR fluid
theory and measure the resulting CR transport coefficients as a function of system parameters.

We implemented this framework in the Athena code, using its MHD-PIC module. In doing so, we
have devised a CR injection recipes at the box boundaries to sustain the CR gradient, and made it
compatible with the $\delta f$ weighting scheme which dramatically reduces the Poisson noise.
We have also presented a theoretical framework for measuring scattering rates from the most
primitive form, to the fluid scattering rate that is weighted average of the primitive rates. We also
derived two versions of theoretically expected scattering rate based on QLT yet with substantially
different predictions, reflecting the dramatic uncertainty and ambiguity in determining such
coefficients in the literature.

We conducted a series of simulations with different resolutions and CR pressure gradients
with ion-neutral damping, and our main findings are as follows.
\begin{itemize}
\item 
At saturated state, wave energy spectrum peaks around the resonant wavelength
of particles with $p\sim p_0$ (characteristic momentum of the CR distribution). The high-$k$ range
develops into a power law wave energy spectrum with $I(k)\propto k^{-1.7}$ following a cascade,
whereas the low-$k$ part of the spectrum develops a different power-law ($\propto k^{0.15}$) before
undergoing a cut-off due to wave damping.

\item
The CRs are not fully isotropized. Particles with momentum $\lesssim$ a few $p_0$ are close to
isotropization away from $\mu=0$, followed by a more steep transition across $\mu=0$. This challenges
the validity of the Eddington approximation with modest reduction of the effective scattering rates.
The Eddington approximation is better applicable to more energetic particles, though most of
these particles are inefficiently scattered due to the cut-off of the low-$k$ wave spectrum.

\item The total fluid scattering rates, after integrating over momentum, conform to the scalings
expected from QLT. In particular, the effective streaming speed is independent of imposed CR
pressure gradient. Compared with theoretical expectations:

-- QLT based on single-fluid treatment of CRs underestimates the effective CR streaming speed
by a factor of $>2$, leading to an overestimate of CR scattering rates by up to an order of magnitude.

-- QLT based on momentum-by-momentum treatment of CRs show reasonable agreement with our
measurements when integrated over momentum near the peak of the distribution function, but bears
large uncertainties at individual momenta especially at low-$p$ ends.

\item When ion-neutral damping dominates, we recommend using (\ref{eq:nuscafluid}) for single-fluid
CR scattering rates with CR energy up to $p=10p_0$, and (\ref{eq:nuscapred}) for
momentum-dependent scattering rates with $p\gtrsim p_0$. A $\nu_{\rm sca}(p)\propto p^{0.7}$ scaling
is applicable in the low-momentum range.

\end{itemize}

Overall, we have provided the first numerical measurement of the CR scattering rates  from
first-principles for the bulk CR population subject to self-confinement, which bridges the gap
between kinetic (CR-gyro) and fluid ($\lambda_{\rm mfp}$) scales. 
We anticipate it will offer realistic sub-grid prescriptions for CR-MHD simulations at
macroscopic scales for studies of CR transport and feedback, thus eliminating a major
source of uncertainty in such simulations. This will help clarify the much unknown role played by
the CRs on scales ranging from the local ISM to those at galactic scales or beyond.

As an initial study, we have limited ourselves to 1D and the simplest ion-neutral damping to demonstrate
the utility of the streaming box framework. This framework is fully generalizable to multi-dimensions,
which can be important to investigate additional scattering mechanisms that are not captured in 1D.
While some of the additional damping mechanisms, such as non-linear Landau damping, requires
resolving the kinetic physics of background plasmas which are missing in our MHD treatment,
such mechanisms can in principle be manually incorporated by operating in Fourier space. Therefore,
we anticipate that our study opens up the window towards fully characterizing the CR scattering rates
over a wide range of regimes, thus providing a comprehensive suite of sub-grid prescriptions for
future CR-MHD simulations of CR transport and feedback.

\begin{acknowledgments}
I thank Eve Ostriker and Chris Bambic for helpful discussions and comments to a
preliminary version of the draft, and the anonymous referee for a thoughtful and
constructive report. This research is supported by the National Science
Foundation of China under Grant No. 11873033.
Numerical simulations are conducted on TianHe-1 (A) at National
Supercomputer Center in Tianjin, China, and on the Orion cluster at
Department of Astronomy, Tsinghua University.
\end{acknowledgments}

\bibliographystyle{apj}

\begin{thebibliography}{86}
\expandafter\ifx\csname natexlab\endcsname\relax\def\natexlab#1{#1}\fi

\bibitem[{{Aloisio} {et~al.}(2015){Aloisio}, {Blasi}, \&
  {Serpico}}]{Aloisio_etal15}
{Aloisio}, R., {Blasi}, P., \& {Serpico}, P.~D. 2015, \aap, 583, A95

\bibitem[{{Amato} \& {Blasi}(2018)}]{AmatoBlasi18}
{Amato}, E. \& {Blasi}, P. 2018, Advances in Space Research, 62, 2731

\bibitem[{{Armillotta} {et~al.}(2021){Armillotta}, {Ostriker}, \&
  {Jiang}}]{Armillotta_etal21}
{Armillotta}, L., {Ostriker}, E.~C., \& {Jiang}, Y.-F. 2021, \apj, 922, 11

\bibitem[{{Bai} {et~al.}(2015){Bai}, {Caprioli}, {Sironi}, \&
  {Spitkovsky}}]{Bai_etal15}
{Bai}, X.-N., {Caprioli}, D., {Sironi}, L., \& {Spitkovsky}, A. 2015, \apj,
  809, 55

\bibitem[{{Bai} {et~al.}(2019){Bai}, {Ostriker}, {Plotnikov}, \&
  {Stone}}]{Bai_etal19}
{Bai}, X.-N., {Ostriker}, E.~C., {Plotnikov}, I., \& {Stone}, J.~M. 2019, \apj,
  876, 60

\bibitem[{{Bambic} {et~al.}(2021){Bambic}, {Bai}, \&
  {Ostriker}}]{Bambic_etal21}
{Bambic}, C.~J., {Bai}, X.-N., \& {Ostriker}, E.~C. 2021, ApJ, in press,
  arXiv:2102.11877

\bibitem[{{Bell}(1978)}]{Bell78}
{Bell}, A.~R. 1978, \mnras, 182, 147

\bibitem[{{Birdsall} \& {Langdon}(2005)}]{BirdsallLangdon05}
{Birdsall}, C.~K. \& {Langdon}, A.~B. 2005, {Plasma Physics Via Computer
  Simulation} (Taylor \& Francis Group, 2005)

\bibitem[{{Blandford} \& {Ostriker}(1978)}]{BlandfordOstriker78}
{Blandford}, R.~D. \& {Ostriker}, J.~P. 1978, \apjl, 221, L29

\bibitem[{{Blasi} {et~al.}(2012){Blasi}, {Amato}, \& {Serpico}}]{Blasi_etal12}
{Blasi}, P., {Amato}, E., \& {Serpico}, P.~D. 2012, \prl, 109, 061101

\bibitem[{{Booth} {et~al.}(2013){Booth}, {Agertz}, {Kravtsov}, \&
  {Gnedin}}]{Booth_etal13}
{Booth}, C.~M., {Agertz}, O., {Kravtsov}, A.~V., \& {Gnedin}, N.~Y. 2013,
  \apjl, 777, L16

\bibitem[{Boris(1970)}]{Boris70}
Boris, J.~P. 1970, in Proceedings of the Fourth Conference on Numerical
  Simulation Plasmas (Navel Research Laboratory, Washington, D.C.), pp. 3--67

\bibitem[{{Breitschwerdt} {et~al.}(1991){Breitschwerdt}, {McKenzie}, \&
  {Voelk}}]{Breitschwerdt_etal91}
{Breitschwerdt}, D., {McKenzie}, J.~F., \& {Voelk}, H.~J. 1991, \aap, 245, 79

\bibitem[{{Chan} {et~al.}(2019){Chan}, {Kere{\v{s}}}, {Hopkins}, {Quataert},
  {Su}, {Hayward}, \& {Faucher-Gigu{\`e}re}}]{Chan_etal19}
{Chan}, T.~K., {Kere{\v{s}}}, D., {Hopkins}, P.~F., {Quataert}, E., {Su},
  K.~Y., {Hayward}, C.~C., \& {Faucher-Gigu{\`e}re}, C.~A. 2019, \mnras, 488,
  3716

\bibitem[{{Cleeves} {et~al.}(2013){Cleeves}, {Adams}, \&
  {Bergin}}]{Cleeves_etal13}
{Cleeves}, L.~I., {Adams}, F.~C., \& {Bergin}, E.~A. 2013, \apj, 772, 5

\bibitem[{{Cohen} \& {Kulsrud}(1974)}]{CohenKulsrud74}
{Cohen}, R.~H. \& {Kulsrud}, R.~M. 1974, Physics of Fluids, 17, 2215

\bibitem[{{Denton} \& {Kotschenreuther}(1995)}]{DentonKotschenreuther95}
{Denton}, R.~E. \& {Kotschenreuther}, M. 1995, Journal of Computational
  Physics, 119, 283

\bibitem[{{Dimits} \& {Lee}(1993)}]{DimitsLee93}
{Dimits}, A.~M. \& {Lee}, W.~W. 1993, Journal of Computational Physics, 107,
  309

\bibitem[{{Drury}(1983)}]{Drury83}
{Drury}, L.~O. 1983, Reports on Progress in Physics, 46, 973

\bibitem[{{Everett} \& {Zweibel}(2011)}]{EverettZweibel11}
{Everett}, J.~E. \& {Zweibel}, E.~G. 2011, \apj, 739, 60

\bibitem[{{Everett} {et~al.}(2008){Everett}, {Zweibel}, {Benjamin}, {McCammon},
  {Rocks}, \& {Gallagher}}]{Everett_etal08}
{Everett}, J.~E., {Zweibel}, E.~G., {Benjamin}, R.~A., {McCammon}, D., {Rocks},
  L., \& {Gallagher}, III, J.~S. 2008, \apj, 674, 258

\bibitem[{{Evoli} {et~al.}(2018){Evoli}, {Blasi}, {Morlino}, \&
  {Aloisio}}]{Evoli_etal18}
{Evoli}, C., {Blasi}, P., {Morlino}, G., \& {Aloisio}, R. 2018, \prl, 121,
  021102

\bibitem[{{Farmer} \& {Goldreich}(2004)}]{FarmerGoldreich04}
{Farmer}, A.~J. \& {Goldreich}, P. 2004, \apj, 604, 671

\bibitem[{{Felice} \& {Kulsrud}(2001)}]{FeliceKulsrud01}
{Felice}, G.~M. \& {Kulsrud}, R.~M. 2001, \apj, 553, 198

\bibitem[{{Ferri{\`e}re}(2001)}]{Ferriere01}
{Ferri{\`e}re}, K.~M. 2001, Reviews of Modern Physics, 73, 1031

\bibitem[{{Foote} \& {Kulsrud}(1979)}]{FooteKulsrud79}
{Foote}, E.~A. \& {Kulsrud}, R.~M. 1979, \apj, 233, 302

\bibitem[{{Ginzburg} \& {Syrovatskii}(1964)}]{GinzburgSyrovatskii64}
{Ginzburg}, V.~L. \& {Syrovatskii}, S.~I. 1964, {The Origin of Cosmic Rays}
  (New York: Macmillan)

\bibitem[{{Girichidis} {et~al.}(2016){Girichidis}, {Naab}, {Walch}, {Hanasz},
  {Mac Low}, {Ostriker}, {Gatto}, {Peters}, {W{\"u}nsch}, {Glover}, {Klessen},
  {Clark}, \& {Baczynski}}]{Girichidis_etal16}
{Girichidis}, P., {Naab}, T., {Walch}, S., {Hanasz}, M., {Mac Low}, M.-M.,
  {Ostriker}, J.~P., {Gatto}, A., {Peters}, T., {W{\"u}nsch}, R., {Glover},
  S.~C.~O., {Klessen}, R.~S., {Clark}, P.~C., \& {Baczynski}, C. 2016, \apjl,
  816, L19

\bibitem[{{Girichidis} {et~al.}(2020){Girichidis}, {Pfrommer}, {Hanasz}, \&
  {Naab}}]{Girichidis_etal20}
{Girichidis}, P., {Pfrommer}, C., {Hanasz}, M., \& {Naab}, T. 2020, \mnras,
  491, 993

\bibitem[{{Girichidis} {et~al.}(2021){Girichidis}, {Pfrommer}, {Pakmor}, \&
  {Springel}}]{Girichidis_etal21}
{Girichidis}, P., {Pfrommer}, C., {Pakmor}, R., \& {Springel}, V. 2021, arXiv
  e-prints, arXiv:2109.13250

\bibitem[{{Grenier} {et~al.}(2015){Grenier}, {Black}, \&
  {Strong}}]{Grenier_etal15}
{Grenier}, I.~A., {Black}, J.~H., \& {Strong}, A.~W. 2015, \araa, 53, 199

\bibitem[{{Guo} \& {Oh}(2008)}]{GuoOh08}
{Guo}, F. \& {Oh}, S.~P. 2008, \mnras, 384, 251

\bibitem[{{Hanasz} {et~al.}(2013){Hanasz}, {Lesch}, {Naab}, {Gawryszczak},
  {Kowalik}, \& {W{\'o}lta{\'n}ski}}]{Hanasz_etal13}
{Hanasz}, M., {Lesch}, H., {Naab}, T., {Gawryszczak}, A., {Kowalik}, K., \&
  {W{\'o}lta{\'n}ski}, D. 2013, \apjl, 777, L38

\bibitem[{{Hawley} {et~al.}(1995){Hawley}, {Gammie}, \& {Balbus}}]{HGB95}
{Hawley}, J.~F., {Gammie}, C.~F., \& {Balbus}, S.~A. 1995, \apj, 440, 742

\bibitem[{{Holcomb} \& {Spitkovsky}(2019)}]{HolcombSpitkovsky19}
{Holcomb}, C. \& {Spitkovsky}, A. 2019, \apj, 882, 3

\bibitem[{{Hopkins} {et~al.}(2021{\natexlab{a}}){Hopkins}, {Chan}, {Ji},
  {Hummels}, {Kere{\v{s}}}, {Quataert}, \&
  {Faucher-Gigu{\`e}re}}]{Hopkins_etal21a}
{Hopkins}, P.~F., {Chan}, T.~K., {Ji}, S., {Hummels}, C.~B., {Kere{\v{s}}}, D.,
  {Quataert}, E., \& {Faucher-Gigu{\`e}re}, C.-A. 2021{\natexlab{a}}, \mnras,
  501, 3640

\bibitem[{{Hopkins} {et~al.}(2021{\natexlab{b}}){Hopkins}, {Squire}, {Chan},
  {Quataert}, {Ji}, {Kere{\v{s}}}, \& {Faucher-Gigu{\`e}re}}]{Hopkins_etal21b}
{Hopkins}, P.~F., {Squire}, J., {Chan}, T.~K., {Quataert}, E., {Ji}, S.,
  {Kere{\v{s}}}, D., \& {Faucher-Gigu{\`e}re}, C.-A. 2021{\natexlab{b}},
  \mnras, 501, 4184
  
 \bibitem[{{Hopkins} {et~al.}(2022{\natexlab{a}}){Hopkins}, {Butsky},
  {Panopoulou}, {Ji}, {Quataert}, {Faucher-Giguere}, \&
  {Keres}}]{Hopkins_etal22a}
{Hopkins}, P.~F., {Butsky}, I.~S., {Panopoulou}, G.~V., {Ji}, S., {Quataert},
  E., {Faucher-Giguere}, C.-A., \& {Keres}, D. 2022{\natexlab{a}},
  arXiv:2109.09762
  
 \bibitem[Hopkins et al.(2022b)]{Hopkins_etal22b}
 Hopkins, P.~F., Squire, J., Butsky, I.~S., et al.\ 2022,
 submitted to \mnras, arXiv:2112.02153

\bibitem[{{Hu} \& {Krommes}(1994)}]{HuKrommes94}
{Hu}, G. \& {Krommes}, J.~A. 1994, Physics of Plasmas, 1, 863

\bibitem[Huang \& Davis(2022)]{HuangDavis22}
Huang, X. \& Davis, S.~W.\ 2022, \mnras doi:10.1093/mnras/stac059

\bibitem[{{Ipavich}(1975)}]{Ipavich75}
{Ipavich}, F.~M. 1975, \apj, 196, 107

\bibitem[{{Ivlev} {et~al.}(2018){Ivlev}, {Dogiel}, {Chernyshov}, {Caselli},
  {Ko}, \& {Cheng}}]{Ivlev_etal18}
{Ivlev}, A.~V., {Dogiel}, V.~A., {Chernyshov}, D.~O., {Caselli}, P., {Ko},
  C.~M., \& {Cheng}, K.~S. 2018, \apj, 855, 23

\bibitem[{{Ji} \& {Hopkins}(2021)}]{JiHopkins22}
{Ji}, S. \& {Hopkins}, P.~F. 2021, arXiv e-prints, arXiv:2111.14704

\bibitem[{{Jiang} \& {Oh}(2018)}]{JiangOh18}
{Jiang}, Y.-F. \& {Oh}, S.~P. 2018, \apj, 854, 5

\bibitem[{{Jokipii}(1966)}]{Jokipii66}
{Jokipii}, J.~R. 1966, \apj, 146, 480

\bibitem[Kempski \& Quataert(2022)]{KempskiQuataert22} 
Kempski, P. \& Quataert, E.\ 2022, submitted to MNRAS, arXiv:2109.10977

\bibitem[{{Kulsrud} \& {Pearce}(1969)}]{KulsrudPearce69}
{Kulsrud}, R. \& {Pearce}, W.~P. 1969, \apj, 156, 445

\bibitem[{{Kulsrud} \& {Cesarsky}(1971)}]{KulsrudCesarsky71}
{Kulsrud}, R.~M. \& {Cesarsky}, C.~J. 1971, \aplett, 8, 189

\bibitem[{{Kunz} {et~al.}(2014){Kunz}, {Stone}, \& {Bai}}]{Kunz_etal14}
{Kunz}, M.~W., {Stone}, J.~M., \& {Bai}, X.-N. 2014, Journal of Computational
  Physics, 259, 154

\bibitem[{{Lazarian}(2016)}]{Lazarian16}
{Lazarian}, A. 2016, \apj, 833, 131

\bibitem[{{Lee} \& {V{\"o}lk}(1973)}]{LeeVolk73}
{Lee}, M.~A. \& {V{\"o}lk}, H.~J. 1973, \apss, 24, 31

\bibitem[{{Mao} \& {Ostriker}(2018)}]{MaoOstriker18}
{Mao}, S.~A. \& {Ostriker}, E.~C. 2018, \apj, 854, 89

\bibitem[{{McKenzie} \& {Voelk}(1982)}]{McKenzieVoelk82}
{McKenzie}, J.~F. \& {Voelk}, H.~J. 1982, \aap, 116, 191

\bibitem[{{Naab} \& {Ostriker}(2017)}]{NaabOstriker17}
{Naab}, T. \& {Ostriker}, J.~P. 2017, \araa, 55, 59

\bibitem[{{Offner} {et~al.}(2019){Offner}, {Gaches}, \&
  {Holdship}}]{Offner_etal19}
{Offner}, S. S.~R., {Gaches}, B. A.~L., \& {Holdship}, J.~R. 2019, \apj, 883,
  121

\bibitem[{{Padovani} {et~al.}(2020){Padovani}, {Ivlev}, {Galli}, {Offner},
  {Indriolo}, {Rodgers-Lee}, {Marcowith}, {Girichidis}, {Bykov}, \&
  {Kruijssen}}]{Padovani_etal20}
{Padovani}, M., {Ivlev}, A.~V., {Galli}, D., {Offner}, S. S.~R., {Indriolo},
  N., {Rodgers-Lee}, D., {Marcowith}, A., {Girichidis}, P., {Bykov}, A.~M., \&
  {Kruijssen}, J.~M.~D. 2020, \ssr, 216, 29

\bibitem[{{Padovani} {et~al.}(2016){Padovani}, {Marcowith}, {Hennebelle}, \&
  {Ferri{\`e}re}}]{Padovani_etal16}
{Padovani}, M., {Marcowith}, A., {Hennebelle}, P., \& {Ferri{\`e}re}, K. 2016,
  \aap, 590, A8

\bibitem[{{Pakmor} {et~al.}(2016){Pakmor}, {Pfrommer}, {Simpson}, \&
  {Springel}}]{Pakmor_etal16}
{Pakmor}, R., {Pfrommer}, C., {Simpson}, C.~M., \& {Springel}, V. 2016, \apjl,
  824, L30

\bibitem[{{Parker} \& {Lee}(1993)}]{ParkerLee93}
{Parker}, S.~E. \& {Lee}, W.~W. 1993, Physics of Fluids B, 5, 77

\bibitem[{{Pfrommer} {et~al.}(2017){Pfrommer}, {Pakmor}, {Schaal}, {Simpson},
  \& {Springel}}]{Pfrommer_etal17}
{Pfrommer}, C., {Pakmor}, R., {Schaal}, K., {Simpson}, C.~M., \& {Springel}, V.
  2017, \mnras, 465, 4500

\bibitem[{{Plotnikov} {et~al.}(2021){Plotnikov}, {Ostriker}, \&
  {Bai}}]{Plotnikov_etal21}
{Plotnikov}, I., {Ostriker}, E.~C., \& {Bai}, X.-N. 2021, \apj, 914, 3

\bibitem[{{Quataert} {et~al.}(2022){Quataert}, {Jiang}, \&
  {Thompson}}]{Quataert_etal21b}
{Quataert}, E., {Jiang}, Y.-F., \& {Thompson}, T.~A. 2022, \mnras, 510, 920

\bibitem[{{Rodgers-Lee} {et~al.}(2020){Rodgers-Lee}, {Taylor}, {Downes}, \&
  {Ray}}]{Rodgers-Lee_etal20}
{Rodgers-Lee}, D., {Taylor}, A.~M., {Downes}, T.~P., \& {Ray}, T.~P. 2020,
  \mnras, 491, 4742

\bibitem[{{Roe}(1981)}]{Roe81}
{Roe}, P.~L. 1981, Journal of Computational Physics, 43, 357

\bibitem[{{Ruszkowski} {et~al.}(2017){Ruszkowski}, {Yang}, \&
  {Zweibel}}]{Ruszkowski_etal17}
{Ruszkowski}, M., {Yang}, H.-Y.~K., \& {Zweibel}, E. 2017, \apj, 834, 208

\bibitem[{{Salem} \& {Bryan}(2014)}]{SalemBryan14}
{Salem}, M. \& {Bryan}, G.~L. 2014, \mnras, 437, 3312

\bibitem[{{Schlickeiser}(2002)}]{Schlickeiser02}
{Schlickeiser}, R. 2002, {Cosmic Ray Astrophysics} (Springer)

\bibitem[{{Schlickeiser} \& {Miller}(1998)}]{SchlickeiserMiller98}
{Schlickeiser}, R. \& {Miller}, J.~A. 1998, \apj, 492, 352

\bibitem[{{Semenov} {et~al.}(2021){Semenov}, {Kravtsov}, \&
  {Caprioli}}]{Semenov_etal20}
{Semenov}, V.~A., {Kravtsov}, A.~V., \& {Caprioli}, D. 2021, \apj, 910, 126

\bibitem[{{Silsbee} \& {Ivlev}(2019)}]{SilsbeeIvlev19}
{Silsbee}, K. \& {Ivlev}, A.~V. 2019, \apj, 879, 14

\bibitem[{{Simpson} {et~al.}(2016){Simpson}, {Pakmor}, {Marinacci}, {Pfrommer},
  {Springel}, {Glover}, {Clark}, \& {Smith}}]{Simpson_etal16}
{Simpson}, C.~M., {Pakmor}, R., {Marinacci}, F., {Pfrommer}, C., {Springel},
  V., {Glover}, S. C.~O., {Clark}, P.~C., \& {Smith}, R.~J. 2016, \apjl, 827,
  L29

\bibitem[{{Skilling}(1971)}]{Skilling71}
{Skilling}, J. 1971, \apj, 170, 265

\bibitem[{{Skilling}(1975)}]{Skilling75a}
---. 1975, \mnras, 172, 557

\bibitem[{{Soler} {et~al.}(2016){Soler}, {Terradas}, {Oliver}, \&
  {Ballester}}]{Soler_etal16}
{Soler}, R., {Terradas}, J., {Oliver}, R., \& {Ballester}, J.~L. 2016, \aap,
  592, A28

\bibitem[{{Squire} {et~al.}(2021){Squire}, {Hopkins}, {Quataert}, \&
  {Kempski}}]{Squire_etal21}
{Squire}, J., {Hopkins}, P.~F., {Quataert}, E., \& {Kempski}, P. 2021, \mnras,
  502, 2630

\bibitem[{{Stone} {et~al.}(2008){Stone}, {Gardiner}, {Teuben}, {Hawley}, \&
  {Simon}}]{Stone_etal08}
{Stone}, J.~M., {Gardiner}, T.~A., {Teuben}, P., {Hawley}, J.~F., \& {Simon},
  J.~B. 2008, \apjs, 178, 137

\bibitem[{{Strong} {et~al.}(2007){Strong}, {Moskalenko}, \&
  {Ptuskin}}]{Strong_etal07}
{Strong}, A.~W., {Moskalenko}, I.~V., \& {Ptuskin}, V.~S. 2007, Annual Review
  of Nuclear and Particle Science, 57, 285

\bibitem[{{Thomas} \& {Pfrommer}(2019)}]{ThomasPfrommer19}
{Thomas}, T. \& {Pfrommer}, C. 2019, \mnras, 485, 2977

\bibitem[{{Thomas} {et~al.}(2021){Thomas}, {Pfrommer}, \&
  {Pakmor}}]{Thomas_etal21}
{Thomas}, T., {Pfrommer}, C., \& {Pakmor}, R. 2021, \mnras, 503, 2242

\bibitem[{{Uhlig} {et~al.}(2012){Uhlig}, {Pfrommer}, {Sharma}, {Nath},
  {En{\ss}lin}, \& {Springel}}]{Uhlig_etal12}
{Uhlig}, M., {Pfrommer}, C., {Sharma}, M., {Nath}, B.~B., {En{\ss}lin}, T.~A.,
  \& {Springel}, V. 2012, \mnras, 423, 2374

\bibitem[{{Wentzel}(1974)}]{Wentzel74}
{Wentzel}, D.~G. 1974, \araa, 12, 71

\bibitem[{{Wiener} {et~al.}(2013){Wiener}, {Oh}, \& {Guo}}]{Wiener_etal13}
{Wiener}, J., {Oh}, S.~P., \& {Guo}, F. 2013, \mnras, 434, 2209

\bibitem[{{Wiener} {et~al.}(2017){Wiener}, {Pfrommer}, \& {Oh}}]{Wiener_etal17}
{Wiener}, J., {Pfrommer}, C., \& {Oh}, S.~P. 2017, \mnras, 467, 906

\bibitem[{{Wiener} {et~al.}(2018){Wiener}, {Zweibel}, \& {Oh}}]{Wiener_etal18}
{Wiener}, J., {Zweibel}, E.~G., \& {Oh}, S.~P. 2018, \mnras, 473, 3095

\bibitem[{{Yan} \& {Lazarian}(2002)}]{YanLazarian02}
{Yan}, H. \& {Lazarian}, A. 2002, Physical Review Letters, 89, 281102

\bibitem[{{Zhao} {et~al.}(2020){Zhao}, {Tomida}, {Hennebelle}, {Tobin},
  {Maury}, {Hirota}, {S{\'a}nchez-Monge}, {Kuiper}, {Rosen}, {Bhandare},
  {Padovani}, \& {Lee}}]{Zhao_etal20}
{Zhao}, B., {Tomida}, K., {Hennebelle}, P., {Tobin}, J.~J., {Maury}, A.,
  {Hirota}, T., {S{\'a}nchez-Monge}, {\'A}., {Kuiper}, R., {Rosen}, A.,
  {Bhandare}, A., {Padovani}, M., \& {Lee}, Y.-N. 2020, \ssr, 216, 43

\bibitem[{{Zirakashvili} {et~al.}(1996){Zirakashvili}, {Breitschwerdt},
  {Ptuskin}, \& {Voelk}}]{Zirakashvili_etal96}
{Zirakashvili}, V.~N., {Breitschwerdt}, D., {Ptuskin}, V.~S., \& {Voelk}, H.~J.
  1996, \aap, 311, 113

\bibitem[{{Zweibel}(2017)}]{Zweibel17}
{Zweibel}, E.~G. 2017, Physics of Plasmas, 24, 055402

\end{thebibliography}

\appendix

\section[]{The Streaming Box Implementation}\label{app:implementation}

\subsection{The Injection Boundary Condition}\label{app:injection}

The key in implementing the streaming box is the CR injection boundary condition.
We first note that the DF of particles {\it entering the box from the boundary}
(i.e. for the CR flux) differs from DF at the boundary, because particles
traveling faster along $\hat{x}$ direction have higher probability to be injected. While it
is possible to work out the new DF, it is not necessarily analytic and
requires additional effort for numerical implementation. Instead, we adopt a much
simpler injection recipes that mimics the real flow of particles and does not require
calculating the DF for CR flux.

Suppose we aim to set up $N$ particles per cell for a CR number density
$n_{\rm CR}^{\rm ctr}$. At the two boundaries, the desired CR number densities are
$n_{\rm CR}^{\rm hi}$ and $n_{\rm CR}^{\rm lo}$ respectively.
Let $\Delta x$ be the cell size, which is uniform, and $\Delta t$ be
simulation timestep. Over time $\Delta t$, the particles can at most travel over a
distance of $\Delta L={\mathbb C}\Delta t$, which is typically comparable to $\Delta x$.
We define
\bgeq
N'={\rm ceil}\bigg(\frac{\Delta L}{\Delta x}N\frac{n_{\rm CR}^{X}}{n_{\rm CR}^{\rm ctr}}\bigg)\ ,\quad
\Delta L'=\frac{N'}{N}\frac{n_{\rm CR}^{\rm ctr}}{n_{\rm CR}^{X}}\Delta  x,
\edeq
where the ${\rm ceil}(y)$ function takes the smallest integer that is larger than $y$,
and superscript `$X$' denotes either ``{\rm lo}" or `{\rm hi}".

In each simulation timestep, we will randomly inject $N'$ particles over a distance of
$\Delta L'$ just {\it outside} of the left/right boundary, 
with each particle $k$ assigned a random location $x_k$, and a random momentum vector
following the desired DF at the boundary. After time
$\Delta t$, these particles would end up at position $x_k'=x_k+v_{kx}\Delta t$. We only keep
the particles whose new positions $x_k'$ are {\it inside} the simulation domain, while other
particles are deleted.

\subsection[]{The $\delta f$ weighting scheme}\label{app:deltaf}

When the CR DF $f({\mb x}, {\mb p})$ is close to some steady background
DF $f_0({\mb x}, {\mb p})$ that is known, one can take this advantage
and consider individual particles as Lagrangian markers representing the difference,
$\delta f$, between $f$ and $f_0$ (instead of $f$). This is handled by a weighting
scheme known as the $\delta f$ method
\citep{ParkerLee93,DimitsLee93,HuKrommes94,DentonKotschenreuther95,Kunz_etal14},
as opposed to the ``full-$f$" scheme.
This $\delta f$ weighting scheme has been implemented to our MHD-PIC code in \citetalias{Bai_etal19},
which dramatically beats down Poisson noise and proves to be indispensable to
accurately follow wave growth and quasi-linear evolution of particles.

The $\delta f$ scheme is based on the Liouville theorem: the DF $f$ is
constant along particle trajectories in phase space. As described in \citetalias{Bai_etal19}, we record the
initial value of $f$ at $t=0$ (i.e., $f_0$) for all particles, and then at every time $t$, a
weight $w_j$ is assigned to each particle $j$ as
\begin{equation}\label{eq:dfweight}
w_j\equiv\frac{\delta f(t, {\mb x}_j(t), {\mb p}_j(t))}{f(t, {\mb x}_j(t), {\mb p}_j(t))}
=1-\frac{f_0({\mb x}_j(t), {\mb p}_j(t))}{f(0, {\mb x}_j(0), {\mb p}_j(0))}\ .
\end{equation}
The CR number density and current density can then be obtained by
\begin{equation}
\begin{split}
\begin{bmatrix}
n_{\rm CR}(t, {\mb x}) \\
{\mb J}_{\rm CR}(t,{\mb x})
\end{bmatrix}
&=\begin{bmatrix}
n_{\rm CR,0}({\mb x}) \\
{\mb J}_{\rm CR,0}({\mb x})
\end{bmatrix}
+\int
\begin{bmatrix}
1 \\
q{\mb v}({\mb p})
\end{bmatrix}
\delta f(t, {\mb x}, {\mb p})d^3{\mb p}\\
&\simeq
\begin{bmatrix}
n_{\rm CR,0}({\mb x}) \\
{\mb J}_{\rm CR,0}({\mb x})
\end{bmatrix}
+
\sum_{j=1}^{N_p}w_j
\begin{bmatrix}
1 \\
q{\mb v}_j
\end{bmatrix}
S({\mb x}-{\mb x}_j)\ ,
\end{split}
\end{equation}
where $S({\mb x})$ is the shape function corresponding to the TSC interpolation scheme,
and $n_{{\rm CR}, 0}$ and ${\mb J}_{{\rm CR}, 0}$ are obtained directly from $f_0$. Note
that the conventional full-$f$ scheme corresponds to setting $w_j=1$, $n_{{\rm CR}, 0}=0$
and ${\mb J}_{{\rm CR}, 0}=0$.

In the streaming box framework, by requiring $\Delta n_{\rm CR}/n_{\rm CR}^{\rm ctr}\ll1$,
it follows that the variation of the CR DF across the box is on the same
order.
Therefore, the advantage of the $\delta f$ weighting scheme is naturally preserved.
The remaining question is, how should we choose the background DF $f_0({\mb x}, {\mb p})$?
We note that $f_0$ does not necessarily be an equilibrium DF, leaving us
some freedom to choose it properly for our convenience, and we can envisage two cases.

\begin{itemize}

\item Case I, suppose CRs are sufficiently scattered so that upon
saturation, the system approaches a quasi-isotropic state with a gradient in accordance
with the imposed CR density gradient. This will yield a background DF
to be
\bgeq
f_0^{\rm I}(x,p)=\frac{n_{\rm CR, 0}(x)}{n_{\rm CR}^{\rm ctr}}f_0^{\rm ctr}(p)\ ,\label{eq:f0case1}
\edeq
where $f_0^{\rm ctr}(p)$ is the mean background (isotropic) DF (or the
desired DF at box center).

\item Case II, consider the opposite limit, where there is no scattering and all CR
particles travel through the box unimpeded. Due to the different CR number density (and
hence flux) imposed at the two boundaries, there is an asymmetry/discontinuity in the overall CR
DF
\bgeq
f_0^{\rm II}(x,p,\mu)=
  \begin{cases}
    [n_{\rm CR}^{\rm hi}/n_{\rm CR}^{\rm ctr}]f_0^{\rm ctr}(p)       & \quad (\mu>0)\\
    [n_{\rm CR}^{\rm lo}/n_{\rm CR}^{\rm ctr}]f_0^{\rm ctr}(p)       & \quad (\mu<0)\\
  \end{cases}\ ,\label{eq:f0case2}
\edeq
where $\mu\equiv\cos\theta$ is the cosine of particle pitch angle (in wave frame, by default),
with $\mu>0$ representing forward traveling CRs.

\end{itemize}

We have experimented both approaches and they yield similar results. In practice,
we choose case I for two reasons. First, our desired saturated state does require
sufficient scattering in the simulation box thus the final CR DF is
closer to (\ref{eq:f0case1}). Second, the initial DF from case I
allows the instability to develop faster. This is because in case I, CR anisotropy
can quickly build up at initial stage (in the absence of initial perturbation) to drive
the CRSI. In case II, however, most of the initial anisotropy lies in the discontinuity
across $90^\circ$ pitch angle, which generally requires finite-amplitude initial
perturbations (which we do not prefer) to ``release" its free energy.

\subsection[]{Choice of frames}\label{app:frame}

In the streaming box framework, additional degrees of freedom arise from the choice
of frames. This choice is two-fold. The first is about the background gas velocity $v_{g0}$
(which is uniform by construction), and the second is related to particle injection, as
one can apply a certain velocity boost $v_{{\rm CR},d}$ to all injected particles.

With periodic boundary conditions for the background gas, different choices of frames are
essentially equivalent. However, we have found that it would be particularly beneficial
to adopt the Alfv\'en wave frame, primarily for numerical reasons:
we find that numerical damping of Alfv\'en waves is much weaker if the simulation is
conducted in the wave frame.
This is helpful given that our simulations must cover a wide range of scales and by working in
the wave frame it leads to effectively higher resolution and hence saves some computational
cost. Therefore, we choose to adopt the wave-frame as a first study.

Once the frame for background gas is settled, choices of different frames for particles
are not equivalent. However, as long as $L\gg\lambda_{\rm mfp}$ and
$v_{g0}, v_{{\rm CR},d}\ll{\mathbb C}$, we anticipate that the CRs should be sufficiently
scattered to establish the desired steady state away from the boundaries. We note that
this state must have certain level of anisotropy as CRs diffuse/stream downward the
pressure gradient, which is a main product of our simulation that we cannot know {\it a priori}.
This means that it is virtually impossible to set the frame of CR injection that perfectly matches
the final level of anisotropy. Therefore, we simply assume zero drift velocity, and inject CRs
isotropically at the boundaries, followed by the injection recipes that we just described.

Because of a small level of mismatch between the injection frame and the frame of CR
streaming in saturated state, there must be a relaxation zone near the boundaries through
which a transition occurs, whose thickness is on the order of $\lambda_{\rm mfp}$.
Therefore, we always conduct measurement near the center of the
simulation box, which is free from boundary artifacts. We also caution that
the CR density/pressure gradient in the central region of the box does not necessarily be
equal to the mean density/pressure gradient that we impose, but has small deviations due to the
boundary relaxation zone. We therefore report results based on the true CR density/pressure
gradients measured in the central region of our simulation box, together with the actual value of
$L_{\rm CR}$ measured in this region.

\section[]{Estimating the CR pitch angle scattering rate}\label{app:derivenu}

\begin{figure}
    \centering
    \includegraphics[width=85mm]{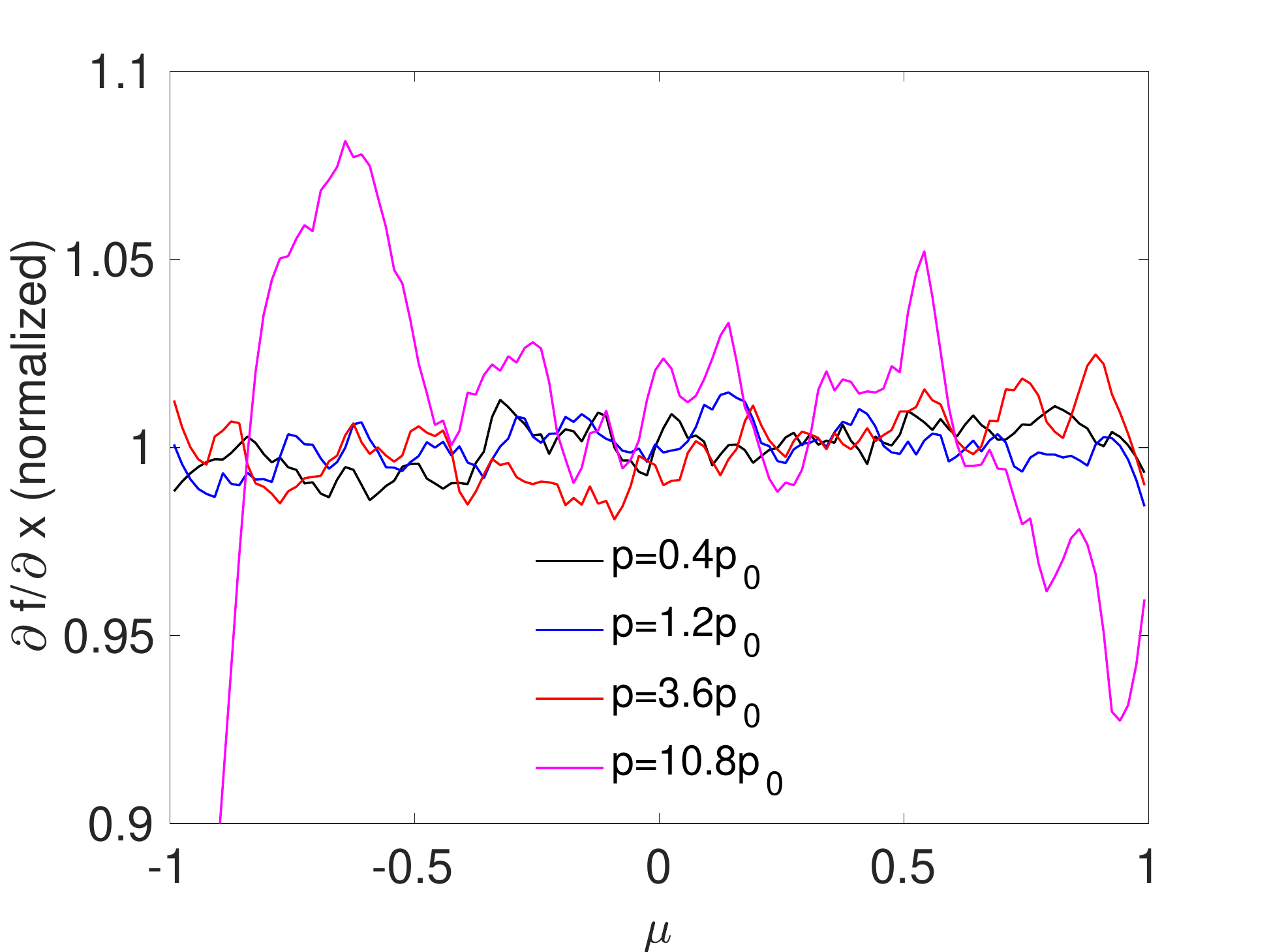}
  \caption{Spatial gradient $\pa f/\pa x$ as a function of $\mu$ for different momentum bins as indicated
  in the legend. The gradient is normalized by $\pa\bar{f}/\pa x$ where $\bar{f}$ is $f$ averaged over
  $\mu$. The lines are largely flat as long as steady state is achieved.}\label{fig:dfdxofmu}
\end{figure}

Here we derive Equation (\ref{eq:nu_ss}) from Equation (\ref{eq:qld}), under the steady state
assumption. In this case, we have
\bgeq\label{eq:qldapp}
\mu V(p)\frac{\pa f}{\pa x}=\frac{\pa}{\pa\mu}\bigg(\frac{1-\mu^2}{2}\nu_\mu(p,\mu)\frac{\pa f}{\pa\mu}\bigg)\ .
\edeq
First of all, we can easily verify that if $\pa f/\pa x$ is independent of $\mu$, then Equation (\ref{eq:nu_ss})
is the exact solution to this equation. As long as $f$ does not strongly deviate from $f_0$, we should
anticipate $\pa f/\pa x\approx\pa f_0/\pa x$, with the latter being independent of $\mu$. Therefore,
we expect Equation (\ref{eq:nu_ss}) to hold at least approximately. That is, $\nu_\mu\pa f/\pa\mu$
should be largely independent of $\mu$. More rigorously, noticing that both $\nu_\mu$ and
$\pa f/\pa\mu$ should be even functions of $\mu$ by symmetry, we may expand it as
\bgeq
\nu_\mu\frac{\pa f}{\pa\mu}\approx A_0+\mu^2A_1+\mu^4A_2+\cdots\ ,
\edeq
with $A_2\ll A_1\ll A_0$, etc. Applying this expansion to (\ref{eq:qldapp}), we obtain
\bgeq
V(p)\frac{\pa f}{\pa x}\approx-(A_0-A_1)-2\mu^2(A_1-A_2)+O(\mu^4)\ .
\edeq
This suggests that $\pa f/\pa x$ must be an even function of $\mu$ as well, which we may write as
$\pa f/\pa x\approx B_0+B_1\mu^2+\cdots$. By matching the terms of the same order, we see
that the error in estimating $\nu_\mu$ from Equation (\ref{eq:nu_ss}) is about
$\delta \nu_\mu/\nu_\mu\sim A_1/A_0\sim B_1/B_0$.

In Figure \ref{fig:dfdxofmu}, we show the ratio of $\pa f/\pa x$ over $\pa\bar{f}/\pa x$, where $\bar{f}$ is the
$\mu$-averaged DF for different particle momenta in our run Fid. It is clear that for particle momenta
$p\lesssim5p_0$, the dependence of $\pa f/\pa x$ on $\mu$ is almost indistinguishable from being flat, which
we estimate $B_1/B_0<1\%$. We already noted that particles with $p\gtrsim10p_0$ has not yet reached steady
state, thus it is conceivable that $\pa f/\pa x$ at these momentum bins show more substantial deviations.
Overall, the results justify the
precision of Equation (\ref{eq:nu_ss}) in estimating the primitive CR pitch angle scattering rate
$\nu_\mu(p,\mu)$.

\section[]{CR momentum feedback and CR heating}\label{app:heating}

As a by-product, the streaming CRs provide momentum and energy feedback to background gas.
Momentum feedback is achieved by momentum exchange between CRs and waves, whose net
effect pushes gas in the direction of CR streaming. Energy feedback is due to wave damping which
heats up the gas, balanced by wave growth driven by the CRSI. This process is better known as
CR heating.

The discussion on the energetics is most convenient in the wave frame, as in our simulations.
From a fluid perspective, momentum conservation
is exhibited as Equation (\ref{eq:fluid1}), where the right hand side reflects the backreaction
to background gas $G(p)$:
\bgeq
G(p)\equiv\frac{\nu_{\rm sca}(p){\cal F}_{\rm CR}}{{\mathbb C}^2}=\frac{4\pi p^3v(p)E(p)}{\mathbb{C}^2}
\int_{-1}^{1}\frac{\pa f}{\pa\mu}\frac{1-\mu^2}{4}\nu(p,\mu)d\mu\ ,
\edeq
and total CR force is given by $\int G(p)d\ln p$.

In the wave frame, there is no energy exchange between the CRs and background gas. Therefore,
growth of wave energy is compensated by reduction of the kinetic energy of the gas,
which can be considered as the work done by $\int G(p)d\ln p$. In steady state, this is balanced by
energy dissipation through wave damping (``CR heating"), and we have
\bgeq\label{eq:heating1}
\Gamma_{\rm heat}=v_A\int G(p)d\ln p=\nu_{\rm in}\langle\rho \delta v^2\rangle
\approx\nu_{\rm in}\rho v_A^2\int I(k)dk=\nu_{\rm in}\rho v_A^2E_{\rm wave}\ ,
\edeq
where angle bracket represents spatial average, and we have taken $\rho\delta v^2\approx\delta B^2$
for Alfv\'en waves.

In the parallel direction, the development of the CRSI accelerates background ions relative to the neutrals,
while ion-neutral drag keeps the ions and neutrals at similar speed. Balancing acceleration with drag, we
have
\bgeq
\int G(p)d\ln p = \nu_{\rm in}\rho\Delta v_g\ ,
\edeq
where $\Delta v_g=v_g-v_{g0}$ is the velocity difference between background ions and neutrals. 
Comparing the above two equations, we obtain
\bgeq\label{eq:deltavg}
\Delta v_g=v_AE_{\rm wave}\ .
\edeq
Therefore, the gas velocity $v_g$ will slightly deviate from our intended velocity of $-v_A$ (so that the
simulation is conducted in the wave frame). Nevertheless, as long as $\Delta v_g\ll v_A$, or
$\delta B^2\ll B_0^2$ (in our simulations on the order of $1\%$ or less), the deviation is largely negligible.
In Table \ref{tab:params}, we can see that given $v_A=1$, $\Delta v_g$ and $E_{\rm wave}$ are largely
identical within $1\%-5\%$, thus confirming the energetics between momentum feedback and CR heating.
Moreover, in the fiducial simulation, the heating rate is about $2.7\times10^{-7}\Omega_c^{-1}$ in code
units, which can can at most contribute to increase temperature by $\lesssim6\%$ over the course of our
simulation. The long heating timescale provides another justification for our use of isothermal equation of
state.

One final piece of verification, as $G(p)$ is balanced by the of CR
pressure gradient from Equation (\ref{eq:fluid1}), we thus anticipate
\bgeq\label{eq:heating2}
\int G(p)d\ln p=\bigg|\frac{\pa P_{\rm CR}}{\pa x}\bigg|\approx\frac{P_{\rm CR}^{\rm ctr}}{L_{\rm CR}}
\approx\nu_{\rm in}\rho\Delta v_g\approx\nu_{\rm in}\rho v_AE_{\rm wave}\ .
\edeq
For the fiducial run, we find $|\pa P_{\rm CR}/\pa x|$, integrated over the entire particle population,
to be $\sim3.04\times10^{-7}$ in code units, higher than the expected value of $\sim2.7\times10^{-7}$.
As we discussed in Section \ref{ssec:primsca}, for particles with $p\lesssim0.1p_0$ and
$p\gtrsim10p_0$, steady state is not quite achieved due to the inefficient scattering (long mean free
paths) and insufficient simulation time for all particles to traverse the simulation box. Therefore, the
gradient of $P_{\rm CR}$ integrated over these momenta is not fully compensated by scattering,
which explains why we find a slightly higher $|\pa P_{\rm CR}/\pa x|$ value.
On the other hand, when integrated over the momentum range of $0.1p_0<p<10p_0$, we obtain the
value of CR pressure gradient to be $2.48\times10^{-7}$. We can interpret this find as particles over
this momentum range accounts for most of the wave excitation and CR scattering.

\label{lastpage}
\end{document}